\documentclass[10pt,twocolumn,twoside]{IEEEtran}
\usepackage{amsmath}
\usepackage{mathtools}
\usepackage{tikz}
\usepackage{verbatim}
\usepackage{subcaption}
\usepackage{pgfplots}
\pgfplotsset{compat=1.10}
\usepackage{graphicx}
\usepackage{epstopdf}
\usepackage{amssymb}
\usepackage{relsize}
\usepackage[textwidth=3.8em,textsize=scriptsize,disable]{todonotes}
\usepackage{algorithm}
\definecolor{Gray}{gray}{0.90}
\usepackage{colortbl}

\usepackage{algorithm}
\usepackage{algorithmicx}
\usepackage{algcompatible}
\usepackage{algpseudocode}

\newtheorem{innercustomgeneric}{\customgenericname}
\providecommand{\customgenericname}{}
\newcommand{\newcustomtheorem}[2]{%
  \newenvironment{#1}[1]
  {%
   \renewcommand\customgenericname{#2}%
   \renewcommand\theinnercustomgeneric{##1}%
   \innercustomgeneric
  }
  {\endinnercustomgeneric}
}

\newcustomtheorem{customthm}{\bf{Theorem}}
\newtheorem{theorem}{\bf{Theorem}}[section]

\newtheorem{cor}[theorem]{\bf{Corollary}}
\newtheorem{lem}[theorem]{\bf{Lemma}}
\newtheorem{prop}[theorem]{\bf{Proposition}}

\newtheorem{remark}[theorem]{\bf{Remark}}
\newenvironment{definition}[1][Definition]{\begin{trivlist}
\item[\hskip \labelsep {\bfseries #1}]}{\end{trivlist}}
\def\qed{\hfill\rule[-1pt]{5pt}{5pt}\par\medskip}
\usepackage{amssymb}
\usepackage{url}
\newcommand{\calG}[0]{\mathcal{G}}
\newcommand{\calV}[0]{\mathcal{V}}
\newcommand{\calE}[0]{\mathcal{E}}

\newcommand{\calA}[0]{\mathcal{A}}
\newcommand{\calB}[0]{\mathcal{B}}

\newcommand{\calN}[0]{\mathcal{N}}

\newcommand{\calD}[0]{\mathcal{D}}
\newcommand{\calM}[0]{\mathcal{M}}

\newcommand{\calL}[0]{\mathcal{L}}
\newcommand{\calK}[0]{\mathcal{K}}
\definecolor{blue}{rgb}{0,0,0.8}
\begin{document}
\title{Trade-off Between Controllability and Robustness in Diffusively Coupled Networks}
\author{Waseem~Abbas, ~\IEEEmembership{Member,~IEEE,}
		Mudassir~Shabbir, 
		 A. Yasin Yaz{\i}c{\i}o\u{g}lu,~\IEEEmembership{Member,~IEEE,} and
        Aqsa~Akber
\thanks{Some preliminary results appeared in \cite{Abbas2019ACC}.}
\thanks{W.~Abbas is with the Electrical Engineering and Computer Science Department at Vanderbilt University, Nashville, TN, USA (e-mail: waseem.abbas@vanderbilt.edu).}
\thanks{M.~Shabbir is with the Computer Science Department at the Information Technology University, Lahore, Punjab, Pakistan (e-mail: mudassir@rutgers.edu).}
\thanks{A. Y.~Yaz{\i}c{\i}o\u{g}lu is with the Department of Electrical and Computer Engineering at the University of Minnesota, Minneapolis, MN, USA (e-mail: ayasin@umn.edu).}
\thanks{A. Akber is with the Electrical Engineering Department at Lahore University of Management Sciences, Punjab, Pakistan (email: 19060048@lums.edu.pk).}
}

\maketitle

\begin{abstract}
In this paper, we demonstrate a conflicting relationship between two crucial properties---\emph{controllability} and \emph{robustness}---in linear dynamical networks of diffusively coupled agents. In particular, for any given number of nodes $N$ and diameter $D$, we identify networks that are maximally robust using the notion of Kirchhoff index and then analyze their strong structural controllability. For this, we compute the minimum number of leaders, which are the nodes directly receiving external control inputs, needed to make such networks controllable under all feasible coupling weights between agents. Then, for any $N$ and $D$, we obtain a sharp upper bound on the minimum number of leaders needed to design strong structurally controllable networks with $N$ nodes and diameter $D$. We also discuss that the bound is best possible for arbitrary $N$ and $D$. Moreover, we construct a family of graphs for any $N$ and $D$ such that the graphs have maximal edge sets (maximal robustness) while being strong structurally controllable with the number of leaders in the proposed sharp bound. We then analyze the robustness of this graph family. The results suggest that optimizing robustness increases the number of leaders needed for strong structural controllability. Our analysis is based on graph-theoretic methods and can be applied to exploit network structure to co-optimize robustness and controllability in networks.
\end{abstract}

\begin{IEEEkeywords}
Network controllability, network robustness, network structure.
\end{IEEEkeywords}

\section{Introduction}
\label{sec:introduction}
\IEEEPARstart{I}{n} a networked control system, controllability and robustness to noise and structural changes in the network are crucial. Controllability describes the ability to manipulate and drive the network to the desired state through external inputs, whereas network robustness expresses the ability of the network to maintain its structure in the event of device or link failures. Another aspect of robustness is the ability to function correctly in the presence of noisy information. Network controllability and robustness are both needed to design networks that achieve desired goals and objectives in practical scenarios. However, it is often observed that networks easier to control exhibit lesser robustness and vice versa, for instance, see \cite{Pasqu2018}. Thus, exploiting trade-offs between network controllability and robustness can have a far-reaching impact on the overall network design.

In this paper, we study the relationship between controllability and robustness in diffusively coupled leader-follower networks represented by undirected graphs. A weighted edge between nodes corresponds to interaction and information exchange between nodes. We consider graphs without self loops. Our focus is on finding extremal networks for the above two properties. In particular, for given parameters such as the number of nodes $N$ and diameter $D$, we consider networks with maximal robustness and then analyze their controllability. Similarly, we design extremal networks that are strong structurally controllable with minimal leaders (input nodes) and then evaluate their robustness. We observe that networks with maximum robustness to noise and structural changes require a large number of control inputs to become controllable, whereas networks that can be controlled through minimum inputs exhibit diminished robustness. In particular, for any given $N$ and $D$ extremal networks for controllability and robustness properties manifest this conflicting behavior.

To characterize network robustness, we utilize a widely used metric \emph{Kirchhoff index} ($K_f$), that captures both aspects of robustness, that is the effect of structural changes in the network as well as the effect of noise on the overall dynamics (for instance, see  \cite{young2010robustness,abbas2012robust,ellens2011effective}). To quantify control performance, we consider the minimum number of inputs (leaders) needed to make the network \emph{strong structurally controllable}, that is, completely controllable irrespective of the coupling weights between nodes (e.g., see  \cite{chapman2013strong,yaziciouglu2016graph,mousavi2018structural}). Accordingly, a network that requires fewer leaders for strong structural controllability is preferred over the one requiring many leaders.

Our approach relies on graph-theoretic methods to exploit the relationship between network controllability and robustness. In \cite{ellens2011effective}, it is shown that for any given number of nodes $N$ and diameter $D$, networks with maximum robustness belong to a particular class of graphs known as \emph{clique chains}. Our main contributions are:
 
\begin{enumerate}
\item For any given $N$ and $D$, we analyze the strong structural controllability of maximally robust graphs, that is, clique chains, and obtain the number of leaders needed for the strong structural controllability of such networks (Section \ref{sec:MRN}). Consequently, we show that for fixed $N$ and $D$, networks with maximal robustness require a large number of control inputs for controllability.

\item For any $N$ and $D$, we obtain a sharp upper bound on the minimum number of leaders that are needed to design strong structurally controllable networks with $N$ nodes and diameter $D$ (Section \ref{sec:MCG}). For this, we utilize the relationship between the dimension of controllable subspace and distances between nodes in a graph, and also discuss that the bound cannot be improved further.

\item We then construct a family of graphs for any $N$ and $D$ such that the graphs are strong structurally controllable with the number of leaders specified in the sharp bound (in point (2) above) and also have the maximal edge sets to achieve the best possible robustness (Section \ref{sec:MCG}).

\item Next, we analyze the robustness of such strong structurally controllable networks. In particular, we provide various upper and lower bounds on Kirchhoff indices of such graphs (Section \ref{sec:Robust_MC}). 

\item Finally, we numerically evaluate our results (Section \ref{sec:num_eval}). Our simulations also indicate that maximally controllable networks are much less robust as compared to clique chains (maximally robust) with the same $N$ and $D$ (Section \ref{sec:robust_sim}).

\end{enumerate}

\subsection{Related Work} Kirchhoff index or equivalently effective graph resistance based measures have been instrumental in quantifying the effect of noise on the expected steady state dispersion in linear dynamical networks, particularly in the ones with the consensus dynamics, for instance, see  \cite{young2010robustness,young2016new,zelazo2017robustness,pirani2017robustness}. Furthermore, limits on robustness measures that quantify expected steady-state dispersion due to external stochastic disturbances in linear dynamical networks are also studied in \cite{siami2016fundamental,siami2017new}. To maximize robustness in networks by minimizing their Kirchhoff indices, various optimization approaches (e.g., \cite{ghosh2008minimizing,summers2015topology}) including graph-theoretic ones \cite{ellens2011effective} have been proposed. The main objective there is to determine crucial edges that need to be added or maintained to maximize robustness under given constraints \cite{mousavi2017robust}.

To quantify network controllability, several approaches have been adapted, including determining the minimum number of inputs (leader nodes) needed to (structurally or strong structurally) control a network, determining the worst-case control energy, and metrics based on controllability Gramians (e.g., see \cite{pasqualetti2014controllability,summers2016submodularity,zhao2019networks}). Since strong structural controllability does not depend on coupling weights between nodes, it is a generalized notion of controllability with practical implications. There have been recent studies providing graph-theoretic characterizations of this concept \cite{chapman2013strong,monshizadeh2014zero,yaziciouglu2016graph,mousavi2018structural}. There are numerous other studies regarding leader selection to optimize various network performance measures under constraints, such as, to minimize the deviation from consensus in a noisy environment \cite{lin2014algorithms,patterson2018resistance}, and to maximize various controllability measures  \cite{yaziciouglu2013leader,olshevsky2015minimum,fitch2016optimal,pequito2017trade,chen2019minimal}. Recently, optimization methods are also presented to select leader nodes that exploit submodularity properties of performance measures for network robustness and structural controllability \cite{summers2016submodularity,clark2017submodularity}.

Recently in \cite{Pasqu2018,pasqualetti2020fragility}, trade-off between controllability and fragility in complex networks is investigated. Fragility measures the smallest perturbation in edge weights to make the network unstable. Pasqualetti \textit{et al.} \cite{Pasqu2018} show that networks that require small control energy, as measured by the eigen values of the controllability Gramian, to move from one state to another are more fragile and vice versa. In our work, for control performance, we consider minimum leaders for strong structural controllability, which is independent of coupling weights; and for robustness, we utilize the Kirchhoff index, which measures robustness to noise as well as to structural changes in the underlying network graph. Moreover, in this work, we focus on designing and comparing extremal networks for these properties.

The rest of the paper is organized as follows: Section \ref{sec:prelim} describes preliminaries, network measures, and also outlines the main problems. Section \ref{sec:Bounds} overviews strong structural controllability bounds in leader-follower networks. Section \ref{sec:MRN} analyzes the controllability of maximally robust networks for given $N$ and $D$. Section \ref{sec:MCN} provides a design of maximally controllable networks and evaluates the robustness of such networks. Section \ref{sec:num_eval} numerically evaluates these results, and finally, Section \ref{sec:con} concludes the paper.
\section{Preliminaries, Network Measures and  Problems}
In this section, we present preliminaries, network controllability and robustness measures, and define our main problems.
\label{sec:prelim}

\subsection{Preliminaries}
\label{sec:sub_prelim}
We consider a network of agents modeled by a simple (loop-free) undirected graph $\calG=(\calV,\calE)$, in which the node set $\calV=\{1,2,\hdots,N\}$ represents agents and the edge set $\calE \subseteq \calV \times \calV$ represents inter-connections between agents. A node $u$ is a neighbor of $v$ if an edge exists between $u$ and $v$, which is denoted by an unordered pair $(u,v)$. The \emph{neighborhood} of $u$ is denoted by $\calN_u = \{v\in\calV | (u,v)\in\calE\}$. The \emph{degree} of node $u$ is simply the number of nodes in its neighborhood, that is $|\calN_u|$. The \emph{distance} between nodes $u$ and $v$, denoted by $d(u,v)$, is the number of edges in the shortest path between $u$ and $v$. The \emph{diameter} of $\calG$, denoted by $\calD$, is the maximum distance between any two nodes in $\calG$. A graph is \emph{weighted} if edges are assigned weights using a \emph{weighting function}
\begin{equation}
\label{eq:weight}
w: \calE\longrightarrow\mathbb{R}^+.
\end{equation}
The \emph{adjacency} matrix of $\calG$ is defined as 
\begin{equation}
\label{eq:adjacency}
\calA_{ij} = 
\left
\lbrace
\begin{array}{cl}
w(i,j) & \text{if } (i,j)\in\calE,\\
0 & \text{otherwise.}
\end{array}
\right.
\end{equation}
Similarly, the \emph{degree} matrix of $\calG$ is defined as 

\begin{equation}
\label{eq:degree}
\Delta_{ij} = 
\left
\lbrace
\begin{array}{cl}
\sum\limits_{k\in\calN_i}\calA_{ik} & \text{if } i=j,\\
0 & \text{otherwise.}
\end{array}
\right.
\end{equation}
The \emph{Laplacian} of $\calG$ is then defined as 
\begin{equation}
\label{eq:laplacian}
\calL = \Delta - \calA.
\end{equation}

We consider that edges in $\calG$ are assigned weights from the interval $[w_{min}, \; w_{max}]$, where $w_{min}$, $w_{max}>0$. However, we assume that the exact values of edge weights are unknown due to system uncertainty. Accordingly, we investigate the network structures with optimal robustness and controllability under the worst-case allocations of edge weights from the feasible set. We provide the corresponding measures of controllability and robustness in the following subsections.
\subsection{Network Controllability Measure}
For the network controllability analysis, we consider a network $\calG = (\calV,\calE)$, in which each agent $i$ updates its state $x_i\in\mathbb{R}$ by the following dynamics
\begin{equation}
\label{eq:consenus}
\dot{x}_i(t) = -\sum\limits_{j\in\calN_i}w(i,j) (x_i(t)-x_j(t)),
\end{equation}
where $w(i,j)$ is the coupling strength between nodes $i$ and $j$.  Moreover, to control and drive the network as desired, external control inputs are injected through a subset of these nodes called \emph{leaders}. The dynamics of the leader node $i$ is,
\begin{equation}
\label{eq:consenus_u}
\dot{x}_i(t) = -\sum\limits_{j\in\calN_i}w(i,j) (x_i(t)-x_j(t)) + u_i(t).
\end{equation}
Let the set of leaders be represented by {$\calV_L=\{\ell_1, \hdots, \ell_k \} \subseteq \calV$}. 
We call the remaining $(\calV \setminus\calV_L)$ nodes that follow the simple consensus dynamics in \eqref{eq:consenus} \emph{followers}. If the total number of nodes is $N$ and the number of leader nodes is $k$, then the overall system level dynamics can be written using the underlying graph's Laplacian as
\begin{equation}
\label{eqn:LTI}
\dot{x}(t) = -\calL x(t) + \calB u(t),
\end{equation}
where $x(t) = \left[\begin{array}{lllll}x_1(t) & x_2(t) & \cdots & x_N(t)\end{array}\right]^T\in\mathbb{R}^N$ be the state vector, $u(t)\in\mathbb{R}^k$ be the control input to the leaders, and $\calB$ be an $N\times k$ input matrix whose $ij^{th}$ entry is 1 if node $i$ is also a leader $\ell_j$, that is
\begin{equation}
\label{eq:B}
\calB_{ij} = \left\lbrace
\begin{array}{cl}
1 & \text{if } i= \ell_j\\
0 & \text{otherwise.}
\end{array}
\right.
\end{equation}

A state $x\in\mathbb{R}^N$ is reachable if there exists some input that can drive the system in (\ref{eqn:LTI}) from origin to $x$ in a finite amount of time. A set of all reachable states constitutes the \emph{controllable subspace}, which is the range space of the following matrix.
\begin{equation}
\label{eqn:C_mat}
\Gamma(\mathcal{G},\mathcal{V}_L) = \left[
\begin{array}{ccccc}
\mathcal{B} & -\mathcal{L}\mathcal{B} & (-\mathcal{L})^2\mathcal{B} & \cdots & (-\mathcal{L})^{N-1}\mathcal{B}
\end{array}
\right]
\end{equation}
Here, $\mathcal{V}_L\subseteq\mathcal{V}$ is the set of leader nodes (defining the input matrix $\mathcal{B}$). The dimension of controllable subspace is the rank of $\Gamma$, which needs to be $N$ for complete controllability. The rank of $\Gamma$ depends not only on the edge set of the graph, but also on the edge weights. In fact, a graph that is controllable for one set of edge weights might not remain controllable if edge weights are changed.

\begin{definition}
\label{def:dimSSC}
For a given graph $\mathcal{G}$ and leader nodes (inputs), the minimum rank of $\Gamma$ for any choice of {non-zero} edge weights is the \emph{dimension of strong structurally controllable} subspace, or simply the dimension of SSC.
\end{definition}

A graph $\mathcal{G}$ is said to be \emph{strong structurally controllable (SSC)} with a given set of leaders $\mathcal{V}_L$, if the resulting controllability matrix $\Gamma(\mathcal{G},\mathcal{V}_L)$ is full rank with \emph{any} choice of {non-zero} edge weights. In this case, we say that $(\mathcal{G},\mathcal{V}_L)$ is a \emph{strong structurally controllable (SSC) pair}. Thus, in a strong structurally controllable network, perturbation in edge weights has no effect on the dimension of controllable subspace, which makes the notion of strong structural controllability particularly useful in situations where precise edge weights are not known due to uncertainties, numerical inaccuracies, and inexact system parameters. As a result, we are interested in finding the {minimum number of leaders required to make a network strong structurally controllable}.

\subsection{Network Robustness Measure}
To analyze network robustness, we utilize the notion of \emph{Kirchhoff index} of a weighted graph, denoted by $K_f(\calG,w)$, and defined as 
\begin{equation}
\label{eqn:K_f}
K_f (\calG,w)= N\sum\limits_{i=2}^N\frac{1}{\lambda_i},
\end{equation} 

where $N$ is the number of nodes and $\lambda_2\le\lambda_3\le\cdots\lambda_N$ are the positive eigenvalues of the weighted Laplacian of the graph as defined in \eqref{eq:laplacian}. 
Our motivation to use $K_f(\calG,w)$ here is twofold. 

First, it is very useful in characterizing the \emph{functional robustness}, which is robustness to noise of linear consensus dynamics over networks. In a connected network $\calG = (\calV,\calE)$, if agents follow the consensus dynamics as in \eqref{eq:consenus}, then global consensus is guaranteed,  ${\lim_{t\to \infty} x(t) \in span \{\bold{1}\}}$, for any $x(0) \in \mathbb{R}^N$. However, in the presence of noise, that is, if {$\dot{x}_i(t)= - \sum_{j\in \mathcal{N}_i}w(i,j) (x_i(t)-x_j(t))+ \xi_i(t)$}, where $\xi(t) \in \mathbb{R}$ is i.i.d. white Gaussian noise with zero mean and unit covariance, perfect consensus cannot be achieved. Instead, some finite steady state variance of $x(t)$ is observed on connected graphs \cite{young2010robustness,bamieh2012coherence}. Accordingly, the robustness of the network to noise is quantified through the expected population variance in steady state, i.e.,

\begin{equation}
\label{heq}
\mathcal{H}(\mathcal{G},w) \coloneqq \lim_{t \to \infty} \frac{1}{N}\sum\limits_{i =1}^{n} \mathrm{E}[{(x_i(t)-\tilde{x}(t))^2}],
\end{equation} 
where $\tilde{x}(t) \in \mathbb{R}$ denotes the average of $x_i(t)$  for all $i$. It is shown in \cite{young2010robustness,yasin2019CDC} that 

\begin{equation}
\label{heq2}
{\mathcal{H}(\mathcal{G},w) = \frac{K_f(\calG,w)}{2N^2}.}
\end{equation}

{Thus, a higher value of $K_f(\calG,w)$ means more dispersion in the steady state, which means the network is less robust to noise and vice versa. In other words, functional robustness and Kirchhoff index are inversely proportional to each other.
}

Second, from a structural viewpoint, Kirchhoff index of a network also captures its \emph{structural robustness}---ability of the network to retain its structural attributes in the case of edge (link) or node deletions. It assimilates the effect of not only the number of paths between nodes, but also their quality as determined by the lengths of paths \cite{ellens2011effective}. For a detailed discussion, we refer the readers to \cite{abbas2012robust,ellens2011effective,ghosh2008minimizing}. {For a higher robustness to noise and structural changes, we desire a network to have a smaller $K_f(\calG,w)$.} Note that for a given $\calG$, $K_f(\calG,w)$ is a function of weights assigned to edges. {In this work, we assess the network  robustness based on the largest possible value of $K_f(\calG,w)$ attained when the edge weights are assigned independently from a bounded interval $[w_{min},\;w_{max}]$ by an adversary, that is,}

\begin{equation}
\label{eq:Kfstar}
\boldsymbol{\calK}(\calG) \coloneqq \max\limits_{\substack{w(i,j)\in[w_{min},\;w_{max}], \\ \forall (i,j)\in\calE}} K_f(\calG,w).
\end{equation}

{In other words, we consider the robustness of network $\calG$ in the worst case with edge weights selected from the interval $[w_{min},\;w_{max}]$, where $w_{min},w_{max}>0$. Since minimizing $\boldsymbol{\calK}(\calG)$ means maximizing the worst-case robustness, our goal for the network design purpose will be to minimize $\boldsymbol{\calK}(\calG)$ and design a network with the maximum worst-case robustness.}

\begin{remark}
{We note that the Kirchhoff index of a weighted graph strictly decreases when edge weights are increased \cite[Theorem~ 2.7]{ellens2011effective}. An immediate consequence is that the solution to the problem in \eqref{eq:Kfstar} is to assign $w_{min}$ to all edges in $\calG$  (as also discussed in \cite{yasin2019CDC}). At the same time, we observe that if all edges in $\calG$ are multiplied by a constant $\alpha>0$, then $K_f(\calG,\alpha w) = K_f(\calG, w)/\alpha$ \cite{yasin2019CDC}. Thus, if we define $K_f(\calG)$ to be the Kirchhoff index of $\calG$ in which all edges have unit weights ($\calG$ is unweighted), then using the above observations, we can write \eqref{eq:Kfstar} as}

\begin{equation}
\label{eq:Kfstar_final}
\boldsymbol{\calK}(\calG) = \frac{K_f(\calG)}{w_{min}}.
\end{equation}

Since $w_{min}$ is some constant, it suffices to consider the Kirchhoff index of the graph with unit weights, that is $K_f(\calG)$, to analyze the network robustness as defined in \eqref{eq:Kfstar}. 
\end{remark}

Thus, from here on, we use $K_f(\calG)$ (Kirchhoff index of the graph with unit edge weights) as the robustness measure of $\calG$. We simply use $K_f$ instead of $K_f(\calG)$ when the context is clear.

\subsection{Problems}
We are interested in exploring relationships between robustness and controllability (as defined above) in diffusively coupled systems (\ref{eqn:LTI}). In particular, we focus on extremal cases, and look at the following problems. 
\begin{itemize}
\item [1.] For given number of nodes $N$ and diameter $D$, maximally robust graphs are clique chains. How many leaders are necessary and sufficient for the strong structural controllability of clique chains?
\item [2.] For any $N$ and $D$, what is the minimum number of leaders needed to design strong structurally controllable networks with $N$ nodes and $D$ diameter?
\item [3.] Construct a family of graphs for any $N$ and $D$ such that graphs have maximal edge sets (for maximal robustness) while being strong structurally controllable with the number of leaders obtained above (in point (2)).
\item [4.] What are the upper and lower bounds on the Kirchhoff index of graphs obtained in point (3)?
\end{itemize}

\section{Background on Strong Structural Controllability and Maximally Robust Networks}
\label{sec:Bounds}
In this section, first, we review tight lower and upper bounds on the dimension of SSC that we will use to compute leaders that are necessary and sufficient for strong structural controllability. Second, we describe networks that are known to have maximum robustness among all networks with $N$ nodes and $D$ diameter. 
\subsection{Lower Bound on the Dimension of SSC based on Distances in Graphs}
\label{subsec:lb}
\begin{figure}[htb]
\centering
\includegraphics[scale=0.6]{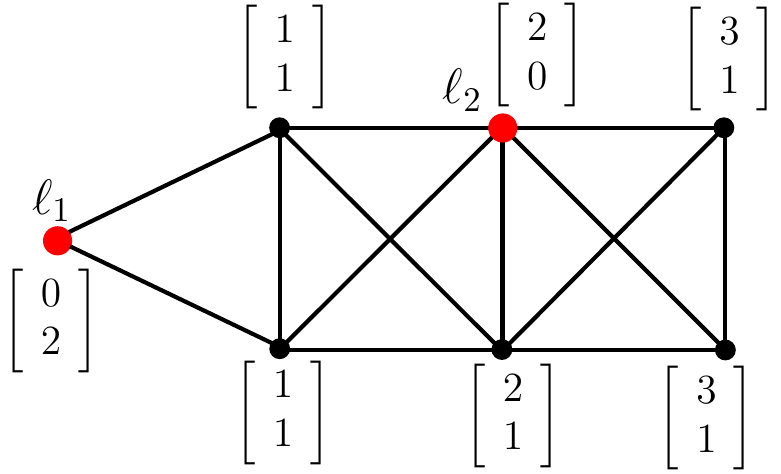}
\caption{A graph with two leaders $\ell_1,\ell_2$ and distance-to-leader vectors of all nodes.}
\label{fig:DL}
\end{figure}

Here, we present a tight lower bound on the dimension of SSC that is based on distances between nodes in a graph \cite{yaziciouglu2016graph}. Let $\calG(\calV,\calE)$ be a leader-follower graph with $k$ leader nodes $\mathcal{V}_L = \{\ell_1,\ell_2,\cdots,\ell_k\}$. For each node $i\in \calV$, we define a \emph{distance-to-leader} vector $S_i\in\mathbb{Z}_+^k$ such that the $j^{th}$ entry of $S_i$, denoted by $S_{i,j}$, is the distance of node $i$ with the leader $j$, that is, 
$$
S_i = \left[ \begin{array}{llllll}d(i,\ell_1) & d(i,\ell_2) & \cdots & d(i,\ell_k)\end{array}\right]^T.
$$
An illustration of the distance-to-leader vectors is shown in Figure \ref{fig:DL}. Next, we construct a sequence of such vectors satisfying some monotonicity conditions.

\begin{definition}{(PMI Sequence)}
\label{def:PMI}
A sequence of distance-to-leader vectors, denoted by $\mathcal{S}=\left[\begin{array}{llllll}S_1 & S_2 & \cdots & S_N\end{array}\right]$, is called a \emph{pseudo-monotonically increasing (PMI)} sequence if for each $S_i\in\mathcal{S}$, there exists an index $\alpha(i)\in\{1,2,\cdots,k\}$ such that 
$$
S_{i,\alpha(i)} \; < S_{j,\alpha(i)}, \;\forall j>i.
$$
\end{definition}
Here, each $S_i\in\mathbb{Z}_+^k$ where $k$ is the number of leaders in the network. We are particularly interested in finding a PMI sequence of distance-to-leader vectors with the maximum length. An example of such a sequence for the network in Figure \ref{fig:DL} is,

$$
\mathcal{S} = \left[
\begin{array}{ccccccc}
\left[
\begin{array}{c}
\textcircled{0} \\ 2
\end{array}
\right], 
\left[
\begin{array}{c}
2 \\ \textcircled{0}
\end{array}
\right], 
\left[
\begin{array}{c}
\textcircled{ 1}\\ 1
\end{array}
\right],
\left[
\begin{array}{c}
\textcircled{2 }\\ 1
\end{array}
\right], 
\left[
\begin{array}{c}
3 \\ \textcircled{1}
\end{array}
\right]
\end{array}
\right].
$$

Note that for each vector, there is an index---of the circled value---such that values of all the subsequent vectors at the corresponding index are strictly greater than the circled value. For instance, the value at the first index is circled in the vector {\scriptsize{$\left[\begin{array}{ccc}0\\2\end{array}\right]$}}, and values at the first indices of all the subsequent vectors are greater than 0. 

We have shown in \cite{yaziciouglu2016graph} that PMI sequences of distance-to-leader vectors in leader-follower networks are particularly useful in studying their strong structural controllability. In this work, we use the following result:

\begin{theorem}\cite{yaziciouglu2016graph}
\label{thm:ssc}
In leader-follower networks, the dimension of SSC is lower bounded by the maximum possible length of any PMI sequence of distance-to-leader vectors of nodes.
\end{theorem}

If the maximum length of a PMI sequence of distance-to-leader vectors in a graph is equal to the number of nodes in a graph, we say that the graph has a \emph{\textbf{full PMI sequence}}. Hence, if $\mathcal{G}$ has a full PMI sequence with a set of leaders $\mathcal{V}_L$, then $(\mathcal{G},\mathcal{V}_L)$ is a strong structurally controllable pair.

\subsection{Upper Bound based on the Dimension of SSC based on the Maximal Leader-Invariant External Equitable Partition}
Here, we discuss an upper bound on the dimension of SSC that is based on a particular partitioning of nodes described as follows: let $\calG(\calV,\calE)$ be a leader-follower network, whose nodes are partitioned into cells $C_1,\cdots,C_k$ such that $\cup_{i=1}^k C_i = \calV$. Let $C_i, C_j$ be two distinct cells and $i\in C_i$, then the \emph{node to cell degree} of $i$ to $C_j$ is $|\mathcal{N}_i\cap C_j|$, and is denoted by $\delta(i,C_j)$. A partition is a \emph{leader-invariant external equitable partition (LIEEP)}, denoted by $\Pi$, if the following conditions are satisfied.
\begin{itemize}
\item[1.] Each leader node is in a singleton cell, that is, if $\ell$ is a leader and it is in a cell $C_\ell$, then $C_\ell = \{\ell\}$.
\item[2.] For any cell $C_i$, let $u,v\in C_i$, then 
$$
\delta(u,C_j) = \delta(v,C_j), \;\forall C_j\ne C_i.
$$  
\end{itemize}
A partition is \emph{maximal LIEEP}, denoted by $\Pi^\ast$, if it is LIEEP and has the minimum number of cells among all LIEEPs. We note that the maximal LIEEP of a graph is unique \cite{zhang2014upper}. 
An important result that relates the notion of maximal LIEEP to controllability in leader-follower networks is as follows:

\begin{theorem}\cite{egerstedt2012interacting,zhang2014upper}
\label{thm:EP}
In an undirected leader-follower network in which each edge has a unit weight, the dimension of controllable subspace is upper bounded by the number of cells in the maximal LIEEP.
\end{theorem}

A direct consequence of the above theorem is that the dimension of SSC in an undirected leader-follower network is upper bounded by the number of cells in the maximal LIEEP. Similarly, we obtain the following corollary.

\begin{cor}
\label{cor:EP}
If $(\mathcal{G},\mathcal{V}_L)$ is a strong structurally controllable pair, then the maximal LIEEP of $\mathcal{G}$ consists of only singleton cells.
\end{cor}

\begin{figure}[htb]
\centering
\includegraphics[scale=0.6]{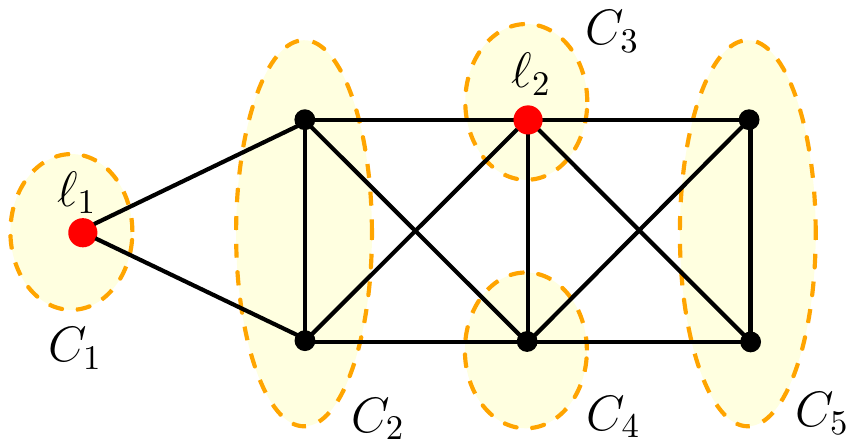}
\caption{Maximal LIEEP of the graph consisting of five cells. $\ell_1$ and $\ell_2$ are leader nodes and are in singleton cells.}
\label{fig:LIEEP}
\end{figure}
\subsection{Maximally Robust Networks}
Next, we describe maximally robust graphs for any fixed $N$ and $D$. These graphs belong to a special class known as the \emph{clique chain}, and have minimum $K_f$ among all graphs with $N$ nodes and $D$ diameter.

\begin{definition}{(\emph{Clique chain \cite{ellens2011effective}}})
Let $n_1,n_2,\cdots,n_D,n_{D+1}$ be a set of positive integers and $N=\sum_{i=1}^{D+1}n_i$, then a clique chain of $N$ nodes and diameter $D$ is a graph obtained from a path graph of diameter $D$, that is $P_{D+1}$, by replacing each node with a clique of size $n_i$ such that the vertices in distinct cliques are adjacent if and only if the corresponding original vertices in the path graph are adjacent. We denote such a clique chain by $\mathcal{G}_D(n_1,\cdots,n_{D+1})$.
\end{definition}

\begin{figure}[htb]
\centering
\includegraphics[scale=0.7]{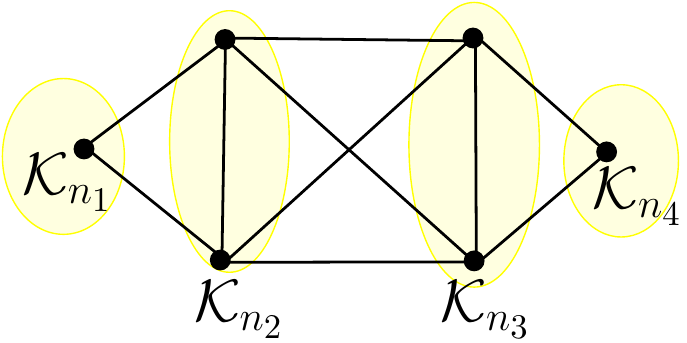}
\caption{A clique chain $\mathcal{G}_3(1,2,2,1)$ with 6 nodes and diameter 3 with $n_1 =1, n_2 = 2, n_3 = 2$, and $n_4=1$.}
\label{fig:clique_chain}
\end{figure}

An example is illustrated in Figure \ref{fig:clique_chain}. 

It is shown in \cite{ellens2011effective} that for given $N$ and $D$, graphs that achieve the minimum $K_f$ are necessarily clique chains of the form ${\mathcal{G}_D(n_1=1,n_2,\cdots,n_D,n_{D+1}=1)}$ where $N = \sum_{i=1}^{D+1}n_i$. Note that the $n_1$ and $n_{D+1}$ are always 1 in optimal clique chains.

\section{{Controllability of Maximally Robust Networks}}
\label{sec:MRN}
In this section, we analyze the strong structural controllability of maximally robust graphs, that is, clique chains. {We show that such networks require a large number of leaders for strong structural controllability.} The main result of this section is stated below.

\begin{theorem}
\label{thm:con_CC}
Let $\mathcal{G}_D(n_1,\cdots,n_{D+1})$ be a clique chain with diameter $D>2$,  and $k$ be the number of leaders needed for the strong structural controllability of $\mathcal{G}_D$, then
\begin{equation}
\label{eqn:bound_CC}
N-(D+1) \; \le \; k \; \le N-D.
\end{equation}
\end{theorem}

We prove this result in Section \ref{sec:proof} by the graph-theoretic tools for the controllability of networked systems. In particular, we utilize the notions of
\begin{itemize}
\item \emph{maximal leader invariant external equitable partitions (LIEEP)} \cite{zhang2014upper,egerstedt2012interacting} to get the lower bound, and
\item the notion of distance-to-leader vectors and \emph{pseudo-monotonically increasing sequences (PMI)} that we introduced in \cite{yaziciouglu2016graph} to get the upper bound. 
\end{itemize}
We have explained these concepts with examples and relevant results in Section \ref{sec:Bounds} for completeness and clarity. 

To obtain the lower bound in (\ref{eqn:bound_CC}), we first note that the maximal LIEEP consisting of only singleton cells is a necessary condition for controllability (Theorem~\ref{thm:EP}). Next, we determine the minimum number of leaders to have such a maximal LIEEP, which directly gives the  minimum number of leaders for strong structural controllability. For the upper bound in (\ref{eqn:bound_CC}), we determine the minimum number of leaders such that the graph has a \emph{full PMI sequence}, which in turn would imply that the network is strong structurally controllable with that many leaders (Theorem~\ref{thm:ssc}). A detailed proof is given below.

\subsection{Proof of Theorem \ref{thm:con_CC}}
\label{sec:proof}
We first prove the lower, and then the upper bound in (\ref{eqn:bound_CC}).
\subsubsection{Lower Bound}
The following result simply states that in the maximal LIEEP of a clique chain, all the non-leader nodes of a clique $\mathcal{K}_{n_i}$ will be in the same cell.
\begin{lem}
\label{lem:CC_2}
Let $\mathcal{G}_D(n_1,\cdots,n_{D+1})$ be a clique chain and $\Pi^\ast$ be its maximal LIEEP. If $u,v$ are non-leader nodes in the same clique $\mathcal{K}_{n_i}$, then they belong to the same cell $C$ of $\Pi^\ast$. 
\end{lem}
\emph{Proof --} Assume $u,v\in \mathcal{K}_{n_i}$ belong to two different cells $C_1$ and $C_2$ of $\Pi^\ast$. Since $u$ and $v$ belong to the same clique, their neighborhoods are exactly the same, which implies $\delta(v,C_x)=\delta(u,C_x)$, $\forall C_x\notin \{C_1,C_2\}$. This means, we can combine $C_1$ and $C_2$ into one cell, and have a LIEEP with one lesser cell, which contradicts that $\Pi^\ast$ is optimal.
\hfill \qed

Next, we show that in the maximal LIEEP of clique chain, a cell that contains non-leader nodes of a clique with a leader(s), contains the non-leader nodes of that clique only.

\begin{lem}
\label{lem:CC_1}
Consider a clique chain $\mathcal{G}_D(n_1,\cdots,n_{D+1})$ with $D>2$. Let $\ell,v$ be respectively, 
a leader and a non-leader node in some clique $\calK_{n_i}$. 
Also let $C_v$ be the cell of $v$ in the maximal LIEEP $\Pi^\ast$ of $G$. For any other node $u\in C_v$, $u$ lies in the same clique $\calK_{n_i}$.
\end{lem}

\noindent
\emph{Proof --} Proof is by contradiction. Let $C_\ell$ be the singleton cell containing $\ell$. Clearly nodes $u,v$ must be neighbors in $\mathcal{G}_D$ as otherwise $\delta(v,C_\ell)\ne \delta(u,C_\ell)$. Assume, without loss of generality, that $u\in \calK_{n_{i+1}}$. Note that $i+1$ is at most $D+1$.

If $i+1<D+1$, let node $w$ belongs to $\calK_{n_{i+2}}$, and be included in a cell $C_w$. Note that $C_w$ cannot contain any node that is adjacent to $\ell$. Since all nodes in the neighborhood of $v$ are adjacent to $\ell$, $C_w$ does not contain any neighbor of $v$. This means that $\delta(v,C_w) = 0$. However, $u$ that is in the same cell as $v$, is adjacent to $w$, and thus has $\delta(u,C_w)>0$, which is not possible in $\Pi^\ast$. Thus $u$ and $v$ are not in the same cell in this case.

If $i+1=D+1$, consider a node $u'\in \calK_{n_{i-1}}$. Since a node $w\in \calK_{n_{i-2}}$ (such a node exists because $D>2$) is adjacent to $u'$ and not adjacent to $v$, $C_{u'}\ne C_v$. By Lemma~\ref{lem:CC_2} all non-leader nodes in $\calK_{n_{i-1}}$ are in $C_{u'}$ and none of the non-leader nodes in $\calK_{n_{i}}\bigcup \calK_{n_{i+1}}$ are in $C_{u'}$. Clearly $\delta(u,C_{u'})< \delta(v,C_{u'})$. Hence, $u$ and $v$ cannot be in the same cell, which is a contradiction.
\hfill \qed

Next we state the following result that directly gives a lower bound in \eqref{eqn:bound_CC}.
\begin{prop}
\label{prop:1}
Let $\mathcal{G}_D(n_1,n_2,\cdots, n_{D+1})$ be a clique chain with $D>2$, then the number of leaders needed to have the maximal LIEEP of $\mathcal{G}_D$ in which each node is in a singleton cell, is at least $N-(D+1)$.
\end{prop}

\emph{Proof --} Let $\Pi^\ast$ be the maximal LIEEP with all nodes in singleton cells. From Lemma \ref{lem:CC_2}, we know that all the non leader nodes of a clique $\mathcal{K}_{n_i}$ will be in the same cell in $\Pi^\ast$. Moreover, from Lemma \ref{lem:CC_1}, we deduce that if $\mathcal{K}_{n_i}$ is a clique with a leader node(s), then all the non-leader nodes of $\mathcal{K}_{n_i}$ will be in the same cell and that cell does not contain a node of any other clique. Thus, we need at least $(n_i-1)$ leaders in the clique $\mathcal{K}_{n_i}$ to have all of its nodes in singleton cells in $\Pi^\ast$. Thus, the minimum number of leaders in $\Pi^\ast$ is $\sum\limits_{i=1}^{D+1} (n_i - 1) = N - (D+1)$. \qed

For strong structural controllability, maximal LIEEP in which each node is in a singleton cell, is a necessary condition (Theorem \ref{thm:EP}). By Proposition \ref{prop:1}, we need at least $N-(D+1)$ leaders to have such a maximal LIEEP, which gives us a lower bound on the number of leaders as in Theorem \ref{thm:con_CC}.

\subsubsection{Upper Bound}
We first state the following result that uses the notion of PMI sequence explained in Section \ref{subsec:lb}.
\begin{lem}
	\label{lem:UB1}
	Let $\mathcal{G}_D(n_1,n_2,\cdots, n_{D+1})$ be a clique chain with $D>2$, then 
	$N-D$ leaders are enough to have a full PMI sequence in $\mathcal{G}_D$.
\end{lem}

\emph{Proof --}
If we add a node $u$ from the first clique to the leader set, then there are at least $D$ nodes (not including $u$) that are at distinct distances from $u$. Save these $D$ nodes, and include all the remaining nodes in the graph to the leader set. With such a set of leader nodes, we get a full PMI sequence of distance-to-leader vectors. \qed

The above lemma implies that $N-D$ leaders are sufficient for the strong structural controllability of clique chains.
\section{Maximally Controllable Networks and their Robustness}
\label{sec:MCN}
In the previous section, we looked at maximally robust networks, and analyzed their controllability. Here, we obtain graphs that are strong structurally controllable with the minimum leaders and evaluate their robustness.
\subsection{Maximally Controllable Networks}
\label{sec:MCG}
For given positive integers $N$ and $D$, let $\mathbb{G}(N,D)$ be the set of all graphs with $N$ nodes and diameter $D$. Moreover, for a given $\calG$, we define $\mathbb{L}(\calG)$ to be the family of all leader sets as follows:

\begin{equation}
\label{eq:BigL}
\mathbb{L}(\calG) := \{\calV_L\subseteq\calV: \; (\calG,\calV_L) \text{ is an SSC pair.} \}
\end{equation}
Then, we define $k_{min}(\calG)$ to be the minimum number of leaders to make $\calG$ strong structurally controllable, that is,

\begin{equation}
\label{eq:kmin}
k_{min}(\calG) := \min\limits_{\calV_L\in\mathbb{L}(\calG)} |\calV_L|.
\end{equation}

We are interested in the minimum value of $k_{min}(\calG)$ among all graphs in $\mathbb{G}(N,D)$. Thus, we define
\begin{equation}
\label{eq:kND}
k_{min}(N,D) := \min\limits_{\calG \in \mathbb{G}(N,D)} k_{min}(\calG).
\end{equation}

Here, our goal will be
\begin{itemize}
\item to compute a sharp upper bound on $k_{min}(N,D)$ (Theorem \ref{thm:graph}), and 
\item to construct graphs with $N$ nodes and $D$ diameter that are strong structurally controllable with the number of leaders specified in the bound.
\end{itemize}

To compute a $k_{min}(N,D)$ bound, we again use the notion of PMI sequences of distance-to-leader vectors. Note that if $\calG$ has a full PMI sequence of distance-to-leader vectors with $\calV_L$ leaders, then $(\calG,\calV_L)$ is an SSC pair.

For a given $\calG$, we define $\mathbb{L}'(\calG)$ to be the family of all leader sets $\calV_L\subseteq\calV$ as follows:
\begin{equation}
\label{eq:BigLp}
\mathbb{L}'(\calG) := \{\calV_L\subseteq\calV: \; \calG \text{ has a full PMI sequence with } \calV_L \}
\end{equation}
Note that $\mathbb{L}'(\calG)\subseteq\mathbb{L}(\calG)$. Then, we define

\begin{equation}
\label{eq:kminp}
{k'_{min}(\calG) := \min\limits_{\calV_L\in\mathbb{L}'(\calG)} |\calV_L|}, 
\end{equation}

and also let
\begin{equation}
\label{eq:kND}
k'_{min}(N,D) := \min\limits_{\calG \in \mathbb{G}(N,D)} k'_{min}(\calG).
\end{equation}

Here, we observe that $k_{min}(\calG) \le k'_{min}(\calG)$, and thus, for any $N$ and $D$
\begin{equation}
\label{eq:kminkminp}
k_{min} (N,D) \le k'_{min}(N,D).
\end{equation}

Next, we focus on computing the exact value of $k'_{min}(N,D)$. For this, we first compute an upper bound on $k'_{min}(\calG)$ for any $\calG$. Recall that \textit{eccentricity} $e_v$ of a node $v$ is defined as the maximum distance of a node from $v$, i.e., $e_v = \max_{u\in V} d(v,u)$. We have the following general result for minimum number of leaders required to have a full PMI sequence:
\begin{theorem}
\label{thm:eccentricity}
Let $\calG$ be a graph with $N$ nodes, and $k$ leaders such that $\calG$ has a full PMI sequence, then 
\[
N\le 1 + \sum_{i=1}^k e_i
\]
where $e_i$ is the eccentricity of leader $\ell_i$.
\end{theorem}

\emph{Proof -- } Without loss of generality, let $[v_1,v_2,\ldots,v_N ]$ be a sequence of nodes whose corresponding distance-to-leader vectors constitute a full PMI sequence in that order. We will construct a sequence of integers (i.e. defined as $s_{(i,N)}$) whose length is same as the full PMI sequence defined above. A bound on length of this sequence will imply the claim of the theorem.
Let $\ell_1\ldots,\ell_k$ be the leader nodes. For a pair of  nonnegative integers $1\le a<b\le N$, we observe that for all leader nodes $\ell_i$, 
\begin{equation}
\label{eqn:Mud_1}
\min_{a\le j\le N} d(\ell_i,v_j) \le \min_{b\le j\le N} d(\ell_i,v_j)
\end{equation} 
Further, by the definition of PMI sequence, there always exists at least one leader $\ell_{i'}$ for which 
\begin{equation}
\label{eqn:Mud_2}
\min_{a\le j\le N} d(\ell_{i'},v_j) < \min_{b\le j\le N} d(\ell_{i'},v_j).
\end{equation}
Next, consider the following sequence of integers,
\begin{equation}
\label{eqn:Mud_seq}
\left[
\begin{array}{llllll}
s_{(1,N)} & s_{(2,N)} & \cdots & s_{(N,N)}
\end{array}
\right],
\end{equation}
where,
\begin{equation}
s_{(a,N)} \triangleq \sum_{i=1}^{k} \min_{a\le j\le N} d(\ell_i,v_j).
\end{equation}

Now, (\ref{eqn:Mud_1}) and (\ref{eqn:Mud_2}) directly imply that the above sequence is a strictly increasing integer sequence with all possible values in the set $\{1,2,\cdots,\sum_{i=1}^k e_i\}\cup \{0\}$, and hence, $N\le 1+\sum_{i=1}^k e_i$ by the pigeonhole-principle.
\qed
Since the maximum eccentricity of a node in a graph is at most the diameter of the graph, Theorem~\ref{thm:eccentricity} provides us the following main result.

\begin{cor}
\label{thm:leaders}
If $\calG\in\mathbb{G}(N,D)$ and $k'_{min}(\calG)$ is the minimum number of leaders needed to have a full PMI sequence of distance-to-leader vectors in $\calG$, then
\begin{equation}
\label{eq:PMI_VL}
k'_{min}(\calG) \ge \left\lceil\frac{N-1}{D}\right\rceil.
\end{equation}
\end{cor}

Next, we show that for any $N$ and $D$, there always exist graphs in $\mathbb{G}(N,D)$ that have full PMI sequences of distance-to-leader vectors (and hence are strong structurally controllable) with exactly $\lceil\frac{N-1}{D}\rceil$ leaders. A direct consequence of this and \eqref{eq:PMI_VL} would be $k'_{min}(N,D) = \lceil\frac{N-1}{D}\rceil$, and then using \eqref{eq:kminkminp}, we would get $k_{min}(N,D) \le \lceil\frac{N-1}{D}\rceil$. To construct such graphs, our approach is as follows:

\begin{itemize}
\item First, for given positive integers $k$ and $D$, we construct a sequence of $N= kD+1$ vectors satisfying the PMI property. Each vector in the sequence is $k$-dimensional and contains values from the set $\{0,1,\cdots,D\}$. 

\item Second, we construct a graph with $N$ nodes and $k$ leaders such that the distance-to-leader vectors of nodes are exactly the same as the vectors obtained in the above step. Thus, the constructed graph has a full PMI sequence of distance-to-leader vectors. The maximum distance between any leader and non-leader node in such a graph will be $D$.

\item Third, we densify the above graph, that is, maximally add edges to the graph while ensuring that the distance-to-leader vectors of nodes do not change. Consequently, we get graphs with $N$ nodes, $D$ diameter and $k$ leaders. Adding edges always reduces $K_f$ and hence, improves robustness \cite{ellens2011effective}. The graphs obtained have full PMI sequences of distance-to-leader vectors, and are strong structurally controllable.
\end{itemize}

To construct sequences, we state the following proposition.

\begin{prop}
Let $S(i,k)$ define the following set of $k$ vectors in $\mathbb{Z}^k$:
$$
S(i,k) =
\left[ 
\begin{matrix}
i &  i+1  & \ldots & i+1\\
i  &  i & \ldots & i+1\\
\vdots & \vdots & \ddots & \vdots\\
i  &   i       &\ldots & i
\end{matrix}
\right],
$$
then the following sequence of $kD+1$ vectors in $\mathbb{Z}^k$ defines a PMI sequence for any positive integers $k$ and $D$.
\begin{equation}
\label{eq:PMI_S}
\left[
\begin{matrix}
0 &  1  & \ldots & 1\\
1  &  0 & \ldots & 1\\
\vdots & \vdots & \ddots & \vdots\\
1  &   1       &\ldots & 0
\end{matrix} \;\; S(1,k)\;\; S(2,k)\;\; \ldots \;\;S(D-1,k)\;\; \begin{array}{c}
D \\ D \\ D \\ \vdots \\ D
\end{array} 
\right]
\end{equation}
\end{prop}

\subsubsection*{Graph Construction} 
\label{sec:graph_con}

Next, we construct a graph $\mathcal{M}$ with $k$ leaders and $N=kD+1$ nodes whose distance-to-leader vectors are same as in \eqref{eq:PMI_S}.
To do so, consider a vertex set 
$$
V = \{\ell_i\} \; \cup \; \{x\} \; \cup \; \{u_{i,j}\}\;,
$$

where $i\in\{1,2\cdots,k\}$ and $j\in\{1,2,\cdots, D-1\}$. Nodes in $\{\ell_1,\ell_2,\cdots,\ell_k\}$ are leaders. We connect these vertices as follows:

\begin{itemize}
\item All leader nodes $\ell_i$ are pair-wise adjacent and induce a clique. 
\item $x$ is adjacent to each $\ell_i$ and $u_{i,1}$, $\forall i\in\{1,\cdots,k\}$.
\item For each $i\in\{2,\cdots,k\}$, $u_{i,1}$ is adjacent to leaders $\ell_{p}$, $ \forall p\in\{i,i+1,\cdots,k\}$.
\item For each $i\in\{1,\cdots,k\}$, $u_{i,j}$ is adjacent to $u_{i,j+1}$, where $j\in\{1,\cdots,D-1\}$.
\end{itemize}

\begin{figure}[htb]
\centering
\includegraphics[scale=0.575]{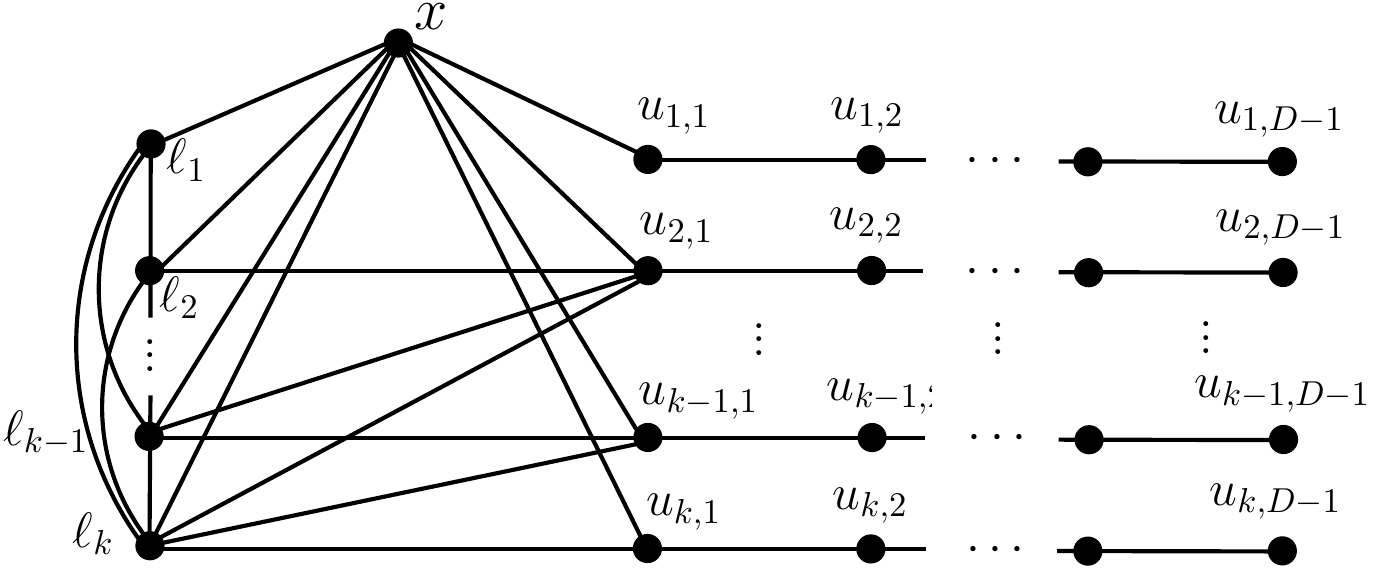}
\caption{Graph $\mathcal{M}$ with $N=kD+1$ nodes, where $k$ is the number of leaders and $D$ is the maximum distance between a leader $\ell_i$ and some other node. Here, $d(\ell_i,u_{1,D-1}) = D, \forall i$.}
\label{fig:construction}
\end{figure}

The above construction is illustrated in Figure \ref{fig:construction}.

Next, we compute the distance-to-leader vectors of nodes in $\mathcal{M}$ as follows:

\begin{itemize}
\item For all $i\in\{1,\cdots,k\}$, the distance-to-leader vector of $\ell_i$ is a vector of all 1's except at the $i^{th}$ index , where it is 0. For the node $x$, it is a vector of all 1's.
\item For node $u_{1,j}$, where $j\in\{1,\cdots,D-1\}$, it is a vector in which all entries are $j+1$.
\item For node $u_{i,j}$, where $i\in\{2,\cdots,k\}$ and $j\in\{1,\cdots,D-1\}$, the distance-to-leader vector has first $(i-1)$ entries equal to $(j+1)$ and the remaining entries are $j$, that is, 

\begin{equation}
\label{eq:DL_vec}
\left[
\begin{array}{ccccccccc}
j+1 & \cdots & j+1 & j & j & \cdots & j
\end{array}
\right]^T.\\
\end{equation}
Here, $j$ is the $i^{th}$ element of the vector.
\end{itemize}

Next, we consider the following sequence of nodes,
\begin{multline}
\label{eqn:seq_nodes}
[ \ell_1, \; \ell_2, \; \cdots, \; \ell_k, \; x, \; u_{2,1}, \;u_{3,1},\; \cdots, \;u_{k,1}, \;u_{1,1},\\
 u_{2,2}, \; u_{3,2}, \; \cdots, \; u_{k,2}, \; u_{1,2},\; u_{2,3}, \; u_{3,3}, \; \cdots, \; u_{k,3}, \; u_{1,3},\\
 \; \cdots, \; 
 u_{2,D-1}, \; u_{3,D-1}, \; \cdots, \; u_{k,D-1}, \; u_{1,D-1}].
\end{multline}

If the distance-to-leader vectors of nodes in $\mathcal{M}$ are arranged in the same order as in \eqref{eqn:seq_nodes}, we get the same sequence as in \eqref{eq:PMI_S}, which is a PMI sequence of length $N$. Hence, $\mathcal{M}$ has a full PMI sequence, and is strong structurally controllable.

\begin{figure}[htb]
\centering
\includegraphics[scale=0.66]{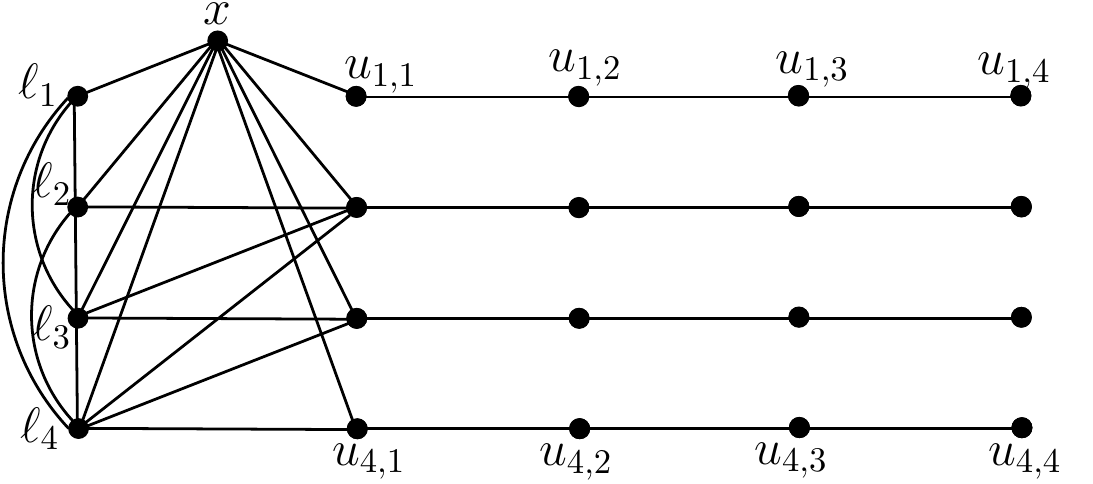}
\caption{A graph $\calM$ with 21 nodes and 4 leaders.}
\label{fig:example_min}
\end{figure}

\emph{Example:} Consider the graph in Figure \ref{fig:example_min}, with $N=21$ nodes and $k=4$ leaders. For any leader $\ell_i$, the maximum distance between $\ell_i$ and any other node is $D = 5$. A full PMI sequence of distance-to-leader vectors is given below. Note that for each vector, there is an index (row index of the circled value) such that the corresponding row value of all the subsequent vectors in the sequence is strictly larger than the circled value, thus constituting a full PMI sequence.

\begin{equation*}
\footnotesize
\left[\begin{array}{llllllllllllllll}
\ell_1 & \ell_2 & \ell_3 & \ell_4 & x & u_{2,1} & u_{3,1} & \cdots & u_{3,4} & u_{4,4} & u_{1,4}\\
\textcircled{0} & 1 & 1 & 1 &  \textcircled{1} & 2 & 2 & & 5 & 5 & \textcircled{5}\\
1   & \textcircled{0}  & 1& 1 & 1 & \textcircled{1} & 2 & \cdots   &  5 & 5 & 5\\
1 & 1 & \textcircled{0} & 1 & 1 & 1 & \textcircled{1} &\cdots  & \textcircled{4} & 5 & 5\\
1 & 1 & 1 & \textcircled{0} & 1 & 1 & 1 &  & 4 & \textcircled{4} & 5\\
\end{array}\right]
\end{equation*}

\subsubsection*{Adding Edges to Graph ${\mathcal{M}}$}
\label{sec:MC}
We note that removing an edge from $\mathcal{M}$ could change the distance-to-leader vectors of nodes. However, we can add edges to $\mathcal{M}$ to improve its robustness by lowering the Kirchhoff index. Next, we construct a new graph $\bar{\mathcal{M}}$ by maximally adding edges to $\mathcal{M}$ while preserving distances between leaders and all other nodes. Consequently, all distance-to-leader vectors and resulting PMI sequence of $\mathcal{M}$ and $\bar{\mathcal{M}}$ are same. We describe the addition of new edges below.

\begin{itemize}
\item For a fixed $j$, all the nodes in $u_{i,j}$, where $i\in\{i,\cdots,k\}$ induce a clique.
\item Each $u_{i,j}$ is adjacent to $u_{1,j-1}$.
\item For a fixed $j>1$, each $u_{i,j}$, where $i>1$, is adjacent to $u_{p,j-1}$, $\forall p\in\{i+1,\cdots,k\}$.
\end{itemize}

\begin{figure}[htb]
\centering
\includegraphics[scale=0.7]{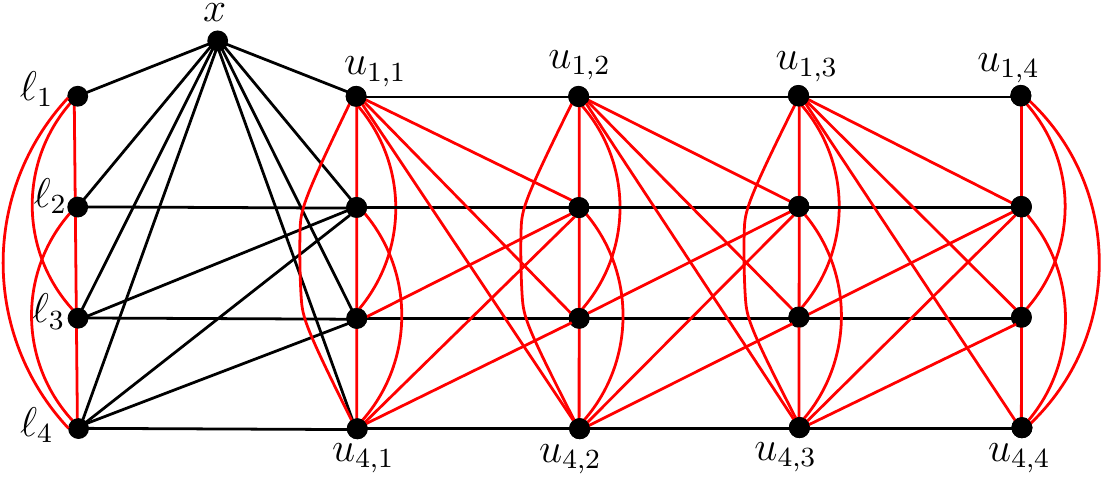}
\caption{Construction of $\bar{\mathcal{M}}$ by adding a maximal edge set (red edges) to $\mathcal{M}$. Here $N=21$, $k=4$ and $D=5$.}
\label{fig:example_max}
\end{figure}

An example of $\bar{\mathcal{M}}$ obtained from $\mathcal{M}$ for $N=21$, $D=5$, and $k=4$ is shown in Figure \ref{fig:example_max}.

\begin{prop}
\label{prop:distance_preservation}
For a fixed $k$ and $D$, the graph $\bar{\mathcal{M}}$ is maximal in the sense that adding any new edge would change the distance-to-leader vector of some node.
\end{prop}
\emph{Proof -- }We classify edges that can be added to $\bar{\mathcal{M}}$ into four types, and will rule them out one by one.
\begin{enumerate}
\item Edge $(x,u_{i,j})$ where $i>1$: such an edge would reduce the distance $d(\ell_1,u_{i,j})$.
\item Edge $(\ell_j,u_{i,j'})$ where $u_{i,j'}\notin \calN_{\ell_j}$: such an edge would reduce the distance $d(\ell_j,u_{i,j'})$.
\item Edge $(u_{1,j},u_{i,j'}) $ where $i>1,j>j'$: such an edge would reduce the distance $d(\ell_{i},u_{1,j})$.
\item Edge $(u_{i,j},u_{i',j'}) $ where $i<i',j<j'$: such an edge would reduce the distance $d(\ell_{i},u_{i',j'})$.
\end{enumerate}
There is only one other edge $(\ell_1,u_{1,1})$, and clearly we cannot add it without changing the distance between $\ell_1$ and $u_{1,1}$.
\qed

Next, we state the following:
\begin{prop}
\label{prop:diameter_M_bar}
If $D$ is the maximum distance between a leader node $\ell_i$ and some other node in $\mathcal{M}$, then $D$ is the diameter of $\bar{\mathcal{M}}$ constructed from $\mathcal{M}$.
\end{prop}

\emph{Proof --} Nodes $u_{1,j},u_{2,j},\ldots, u_{k,j}$ make a clique for all ${1\le j\le D-1}$, and ${u_{i,1},u_{i,2},\ldots,u_{i,D-1}}$ is a path of length $D-2$. Therefore ${d(u_{i,j},u_{i',j'})\le D}$ for all such pairs of nodes. Since all distance-to-leader vectors are preserved in $\bar{\mathcal{M}}$ due to Proposition~\ref{prop:distance_preservation}, farthest node from each leader is still at distance $D$. Thus the graph $\bar{\mathcal{M}}$ has diameter $D$. \qed

\noindent
\textbf{Remark 1 --} So far, we have assumed that $N=kD+1$ for some integer $k$. However, we can obtain the desired graph for any $N$ by modifying $\bar{\mathcal{M}}$. Let $N_a$ be the actual number of nodes, and $D$ be the desired diameter, then we construct a graph $\bar{\mathcal{M}}$ with $N =kD+1$ nodes where $k=\lceil\frac{N_a-1}{D}\rceil$. We need at least that many leaders to have a graph with a full PMI sequence (Theorem \ref{thm:leaders}). Since $N_a<N$, we need to delete $(N-N_a)$ nodes from $\bar{\mathcal{M}}$. We delete the required number of nodes in the following order: first, we delete the nodes (in the same order) $u_{1,D-1}, u_{k,D-1}, u_{k-1,D-1},u_{k-2,D-1}, \cdots, u_{3,D-1}$, then $u_{1,D-2}, u_{k,D-2}, u_{k-1,D-2}, u_{k-2,D-2},\cdots, u_{3,D-2}$, and so on until the total number of nodes in the remaining graph is $N_a$. Note that the nodes $u_{2,D-i}$, where $i\in\{1,2,\cdots\}$ are not deleted to preserve the diameter $D$. In fact, it is easy to verify that as a result of nodes deletion, the distance-to-leader vectors of nodes in the remaining graph remain the same as in the original graph, and hence the maxium length PMI sequence of distance-to-leader vectors of nodes in the remaining graph has length $N_a$ (full PMI sequence). Thus, we can state the following proposition.

\begin{prop}
\label{prop:gen_cons}
For any $N$ and $D$, there exist graphs in $\mathbb{G}(N,D)$ that have full PMI sequences of distance-to-leader vectors with $k = \lceil\frac{N-1}{D}\rceil$ leaders.
\end{prop}

Now, we can state one of the main results of this section.

\begin{theorem}
\label{thm:graph}
For any positive integers $N$ and $D$, 
\begin{equation}
\label{eqn:tight_bnd}
k_{min}(N,D) \le \left\lceil\frac{N-1}{D}\right\rceil.    
\end{equation}
\end{theorem}
\emph{Proof -- } Since having full PMI sequences is a sufficient condition for strong structural controllability (Theorem \ref{thm:ssc}), and since we can construct graphs with full PMI sequences of distance-to-leader vectors for any $N$ and $D$ with $k=\lceil\frac{N-1}{D}\rceil$ leaders (Proposition \ref{prop:gen_cons}), we get the desired result as a direct consequence. \qed

\textbf{Remark 2 --}  The above bound on the number of leaders is tight and cannot be improved for arbitrary $N$ and $D$. In other words, there are graph classes for which we need at least $\lceil\frac{N-1}{D}\rceil$ leaders for strong structural controllability, for instance path graphs ($D=N-1$ and $k=1$), cycle graphs ($D=\lceil N/2 \rceil$ and $k=2$), complete graphs ($D = 1$ and $k=N-1$).

\textbf{Remark 3 --} For any $N$ and $D$, we explicitly define a family of graphs $\bar{\calM}$ using the above construction that achieve the bound in Theorem \ref{thm:graph} and have the maximal edge set. Whether  there exist other families of graphs that achieve the bound in Theorem \ref{thm:graph} and possibly have better Kirchoff index than the graphs in $\bar{\calM}$ remains as an open problem. 
\subsection{Robustness of Maximally Controllable Networks}
\label{sec:Robust_MC}
\begin{figure}[h]
\centering
\includegraphics[scale=0.75]{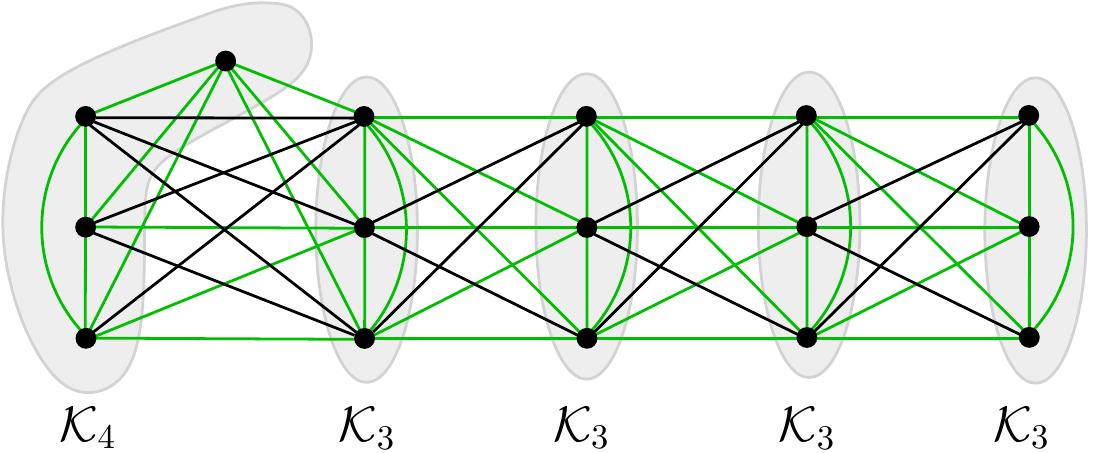}
\caption{A clique chain $\mathcal{G}_4(4,3,3,3,3)$ that contains a maximally controllable graph $\bar{\mathcal{M}}$ with 16 nodes and diameter 5 as a subgraph. Edges in $\bar{\mathcal{M}}$ are highlighted in green.}
\label{fig:CCvsOurs}
\end{figure}

In this subsection, we analyze the robustness of maximally controllable graphs $\bar{\calM}$ by providing bounds on their $K_f$. In the next section, we compare the robustness of $\bar{\calM}$ and clique chains with the same $N$ and $D$. We provide a pair of lower bounds on $K_f(\bar{\calM})$ in Lemmas \ref{lem:M_bound} and \ref{lem:2nd_bound}, and an upper bound on $K_f(\bar{\calM})$ in Lemma \ref{lem:upper_bound}. 
\begin{lem}
\label{lem:M_bound}
Let $k>2$ be a positive integer, $D$ be the diameter, and $N=kD + 1$ be the total number of nodes in a maximally controllable graph $\bar{\mathcal{M}}$ in Section \ref{sec:MCG}, then
\begin{equation}
\label{eq:KfMCG}
K_f(\bar{\mathcal{M}}) > \frac{D^3}{6} + \frac{(k-1)}{3}D^2 + \frac{(k-2)}{3} D.
\end{equation}
\end{lem}
\emph{Proof --} We observe that for a given $D,k$ and $N$, the graph $\bar{\mathcal{M}}$ is a subgraph of a clique chain of the form $\mathcal{G}_{D-1}(k+1,k,k\cdots,k)$ (as illustrated through an example in Figure~\ref{fig:CCvsOurs}). The diameter of this clique chain is $D-1$. Since the Kirchhoff index of a graph is strictly lesser than the Kirchhoff index of any of its proper subgraph, 
\begin{equation*}
K_f(\bar{\mathcal{M}}) > K_f\left(\mathcal{G}_{D-1}(k+1,k,k\cdots,k)\right).
\end{equation*}
From a closed form expression for Kirchhoff Index for a clique chain in \cite[eq. (13)]{ellens2011effective}, we have, 
{\small
\begin{equation*}
\begin{split}
K_f\left(\mathcal{G}_{D-1}(k+1,k,k\cdots,k)\right) =
\frac{D^3}{6} + 
\left(\frac{k^2 -k + 3/2}{3k}\right)D^2\\
+ \left(\frac{12k^4 - (15k^3 + 75 k^2 + 55k +11)}{6k(3k+1)(2k+1)}+\frac{2}{3}\right)D \\
+ \left(\frac{10k^3+18k^2+13k+3}{2k(3k+1)(2k+1)}- \frac{1}{2}\right).
\end{split}
\end{equation*}
}
After simplification and ignoring lower order terms, we have the following:
\begin{equation*}
\begin{split}
    K_f(\bar{\mathcal{M}}) & > K_f\left(\mathcal{G}_{D-1}(k+1,k,k\cdots,k)\right)  \\
    & > \frac{D^3}{6} + \frac{(k-1)}{3}D^2 + \frac{(k-2)}{3} D.
\end{split}
\end{equation*}
\qed
As an example, consider $k=3$, $D=5$, and $N = 16$. A clique chain $\mathcal{G}_4(4,3,3,3,3)$ of diameter 4 is shown in Figure~\ref{fig:CCvsOurs}. Note that $\bar{\mathcal{M}}$ with a diameter 5 and consisting of 16 nodes is a subgraph of $\mathcal{G}_4(4,3,3,3,3)$.

\begin{lem}
\label{lem:2nd_bound}
Let $k$ be a positive integer, and $\bar{\mathcal{M}}$ be the maximally controllable graph with diameter $D$ and $N=kD+1$ nodes, then
\begin{equation}
\label{eq:deg_bound}
K_f(\bar{\mathcal{M}}) > \frac{D^2k(Dk+1)}{2Dk-k+1}.
\end{equation}
\end{lem}

\emph{Proof --} Let $deg(v)$ denote the degree of node $v$ in $\calG$. It is shown in \cite{mieghem_2010,yasin2019CDC} that the Kirchhoff index of any connected graph $\calG$ with $N$ nodes is lower bounded by $(N-1)^2/deg_{av}(\calG)$, where $deg_{av}(\calG) = (1/|\calV|)\sum\limits_{v\in\calV}deg(v)$ is the average degree. It can be shown that the average degree of a maximally controllable graph $\bar{\mathcal{M}}$ with diameter $D$ and $N=kD+1$ nodes is

$$
deg_{av}(\bar{\mathcal{M}}) = \frac{k^2(2D-1) + k}{kD + 1}.
$$
Since $K_f(\bar{\mathcal{M}}) > (N-1)^2/deg_{av}(\bar{\mathcal{M}})$, the desired result follows directly.

\qed

\begin{figure}[h]
\centering
\begin{subfigure}{0.24\textwidth}
\centering
\includegraphics[scale=0.22]{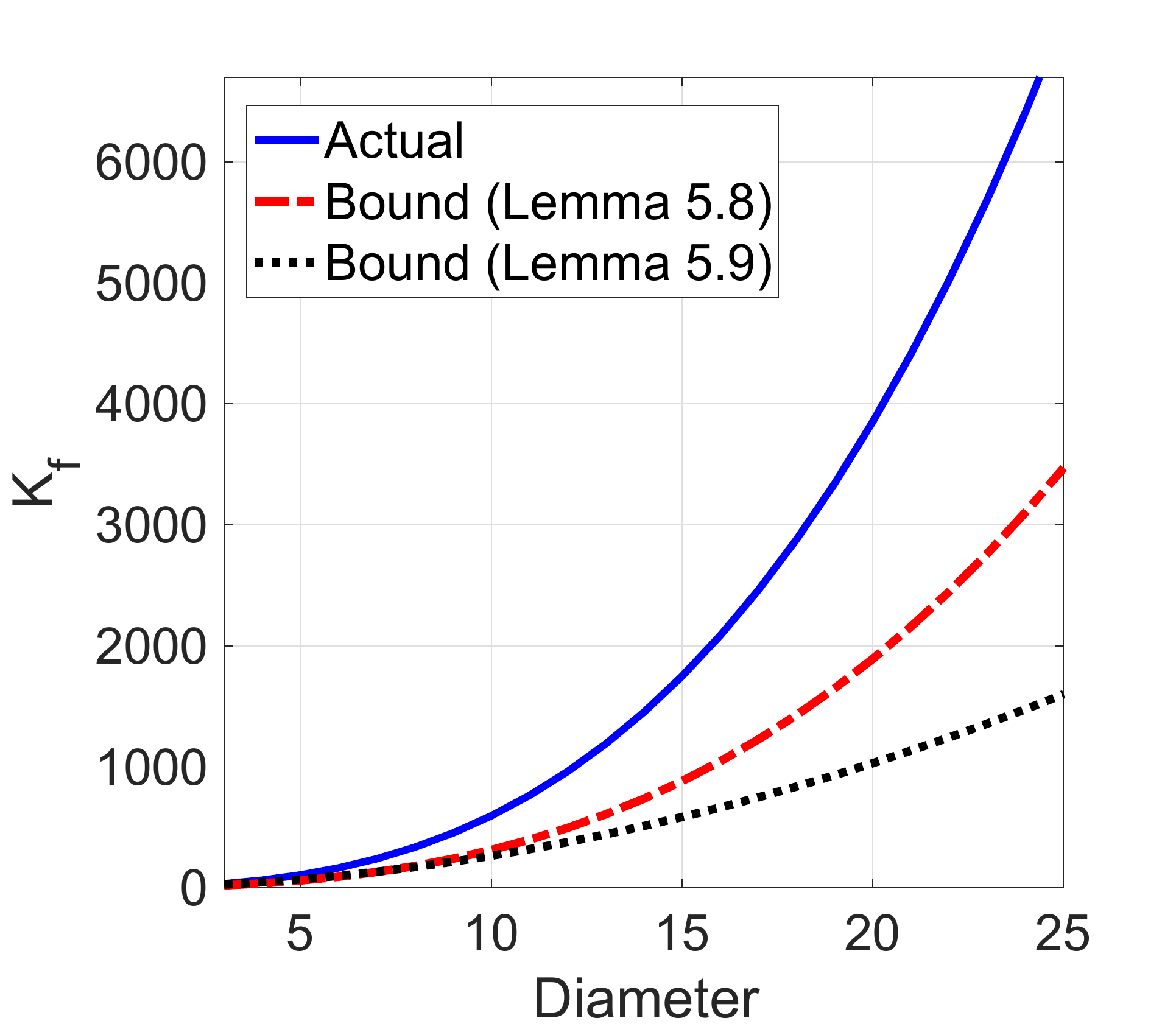}
\caption{$k = 5$}
\end{subfigure}
\begin{subfigure}{0.22\textwidth}
\centering
\includegraphics[scale=0.22]{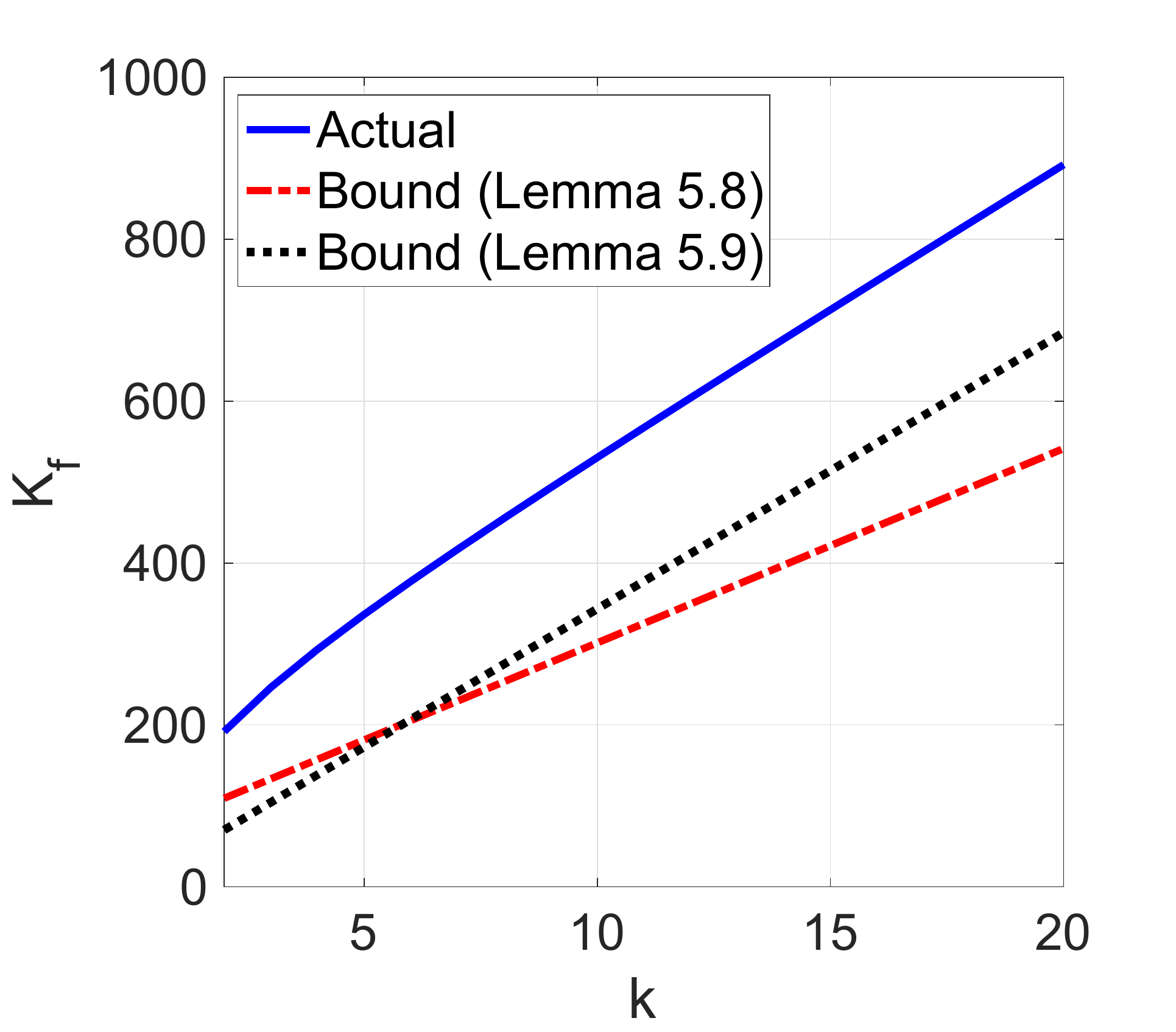}
\caption{$D = 8$}
\end{subfigure}
\caption{Comparison of $K_f$ of the maximally controllable graphs $\bar{\mathcal{M}}$ and the corresponding optimal clique chains with the same $D$ and $N = kD+1$.}
\label{fig:bounds}
\end{figure}

We note that both lower bounds (\eqref{eq:KfMCG} and \eqref{eq:deg_bound}) complement each other for different values of $k$ and $D$. We illustrate this in Figure \ref{fig:bounds}, in which for larger $D$, the bound in \eqref{eq:KfMCG} is better, whereas, for larger $k$, \eqref{eq:deg_bound} is better. So, for any $D$ and $k$, we can simply select the larger of \eqref{eq:KfMCG} and \eqref{eq:deg_bound} as a lower bound on $K_f(\bar{\mathcal{M}})$.

\begin{lem}
\label{lem:upper_bound}
Let $k$ be a positive integer, and $\bar{\mathcal{M}}$ be the maximally controllable graph with diameter $D$, and $N=kD+~1$ nodes, then
\begin{equation}
\label{eq:dis_upper_bound}
\begin{split}
K_f(\bar{\mathcal{M}})  <   k\binom{D}{2}\left[\frac{1}{2} + k + \frac{(2D-1)k}{6}\right] + D\dbinom{k+1}{2}.\\
\end{split}
\end{equation}
\end{lem}

\emph{Proof --}
It has been shown in \cite{yasin2019CDC} that
\begin{equation}
\label{eq:KfUpperGen}
K_f(\calG) \le \frac{N(N-1)}{2} dis_{av}(\calG),
\end{equation}
where $dis_{av}(\calG)$ denotes the average distance in $\calG$, that is, 
\begin{equation}
\label{eq:avdis}
dis_{av}(\calG) = \frac{2 \left(\sum \limits_{1\le i<j\le N} d(i,j)\right)}{N(N-1)}, 
\end{equation}
and equality holds in \eqref{eq:KfUpperGen} if and only if $\calG$ is a tree. Thus, from \eqref{eq:KfUpperGen} and \eqref{eq:avdis}, we have 
\begin{equation}
\label{eq:UpperProof}
K_f(\bar{\calM}) < \sum_{1\le i<j\le N} d(i,j).
\end{equation}
For any given $\bar{\calM}$, computation of all pair-wise distances between nodes and their summation gives the following:
\begin{equation}
\label{eq:MdisSum}
\sum_{1\le i<j\le N} d(i,j) = k\binom{D}{2}\left[\frac{1}{2} + k + \frac{(2D-1)k}{6}\right] + D\dbinom{k+1}{2}.
\end{equation}
The desired result follows directly from \eqref{eq:UpperProof} and \eqref{eq:MdisSum} \qed.

\section{Numerical Evaluation}
\label{sec:num_eval}
In this section, we numerically evaluate our results by comparing controllability and robustness of clique chains and maximally controllable networks with the same $N$ and $D$.

\subsection{Controllability Comparison}
\begin{figure}[h]
\centering
\begin{subfigure}{0.24\textwidth}
\centering
\includegraphics[scale=0.3]{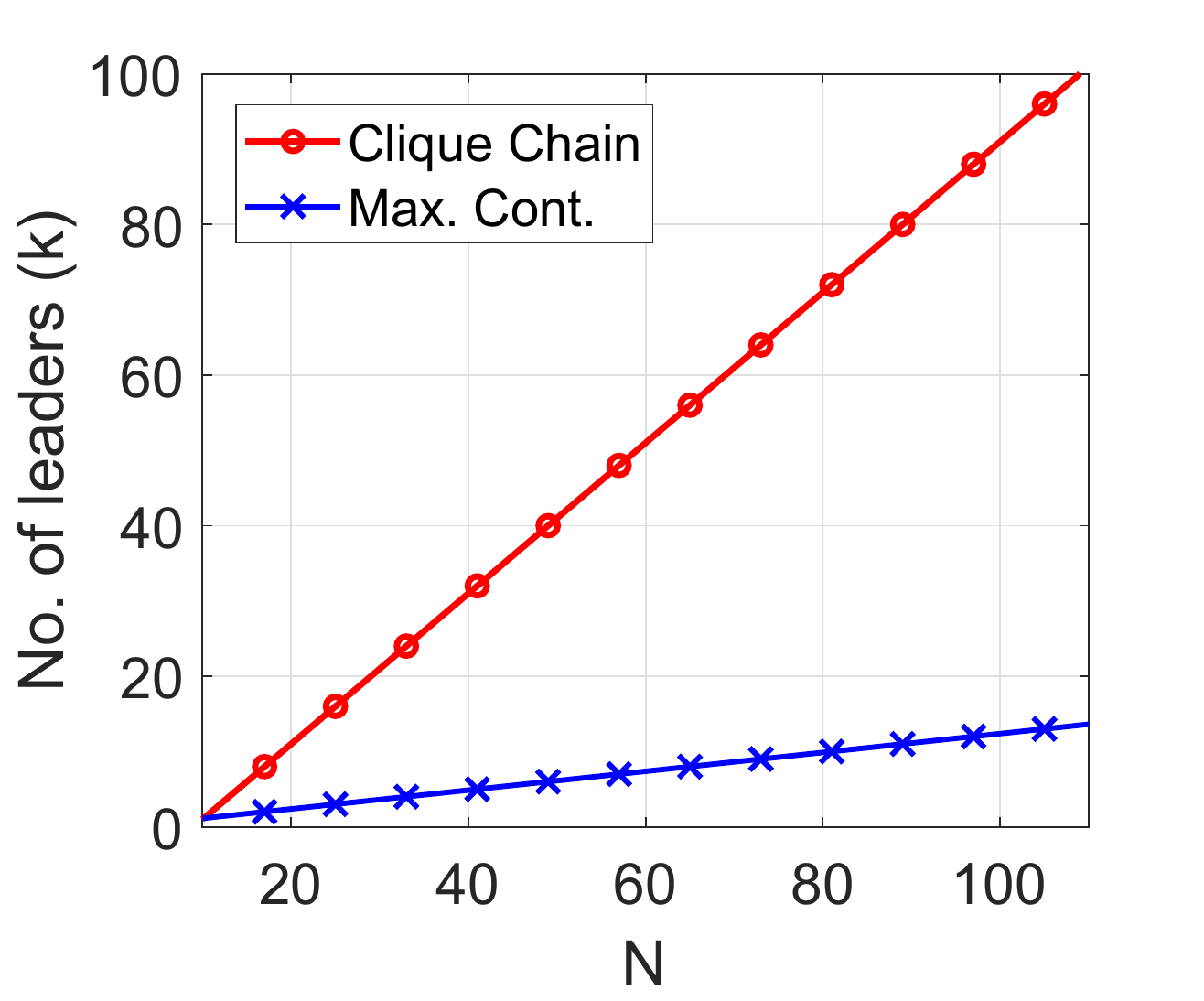}
\caption{$D=8$}
\end{subfigure}
\begin{subfigure}{0.22\textwidth}
\centering
\includegraphics[scale=0.3]{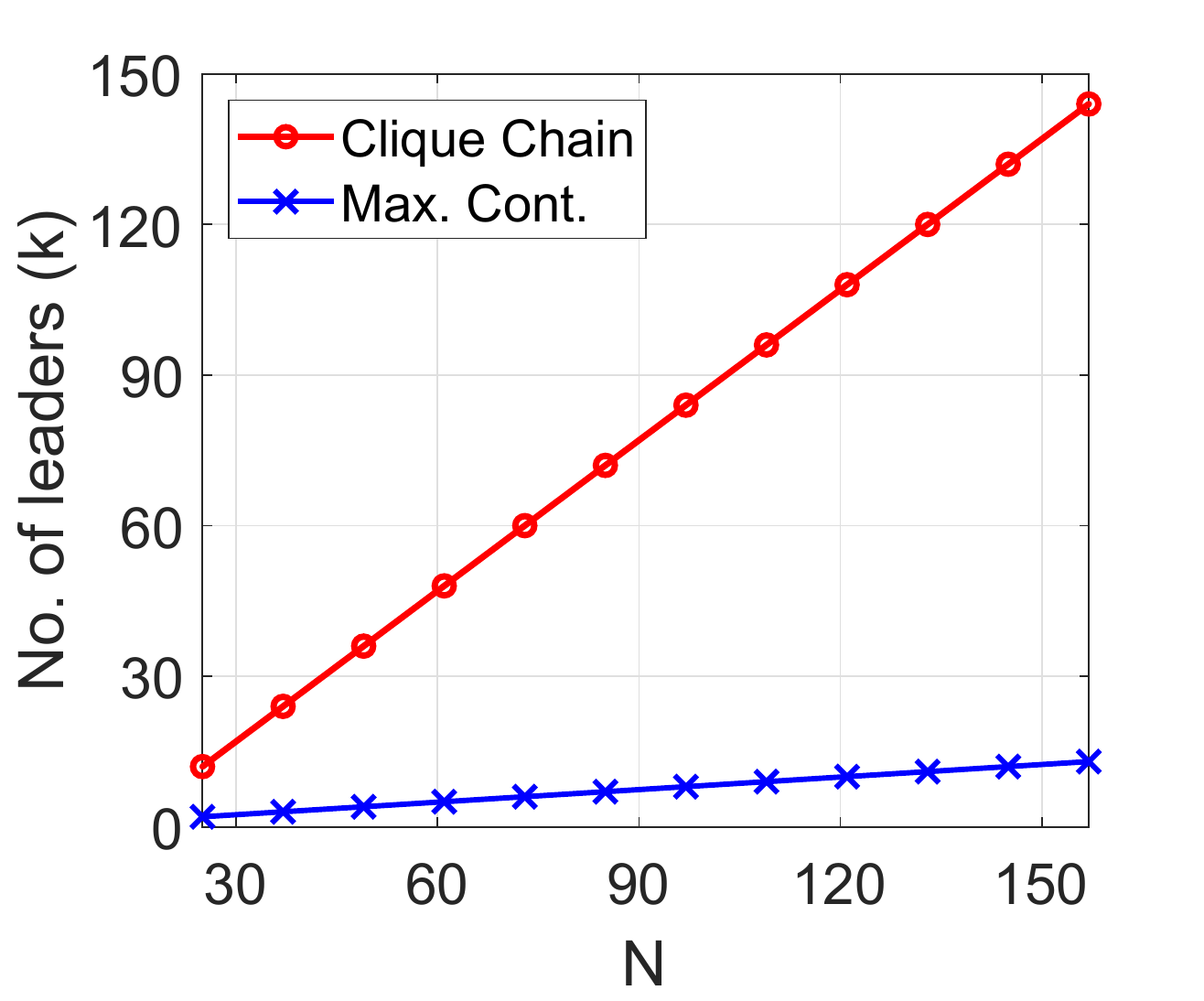}
\caption{$D=12$}
\end{subfigure}
\caption{Comparison of the number of leaders for the strong structural controllability of clique chains and corresponding maximally controllable graphs as a function of $N$.}
\label{fig:con_comp}
\end{figure}

We illustrate the the number of leaders needed for the strong structural controllability of maximally robust networks, that is clique chains for given $N$ and $D$ using the lower bound in \eqref{eqn:bound_CC}. Theorem \ref{thm:graph} states that for given $N$ and $D$ there always exists a graph that is strong structurally controllable with $\lceil\frac{N-1}{D}\rceil$ leaders, such as the maximally controllable graph $\bar{\calM}$ constructed in Section \ref{sec:MCG}. For both graphs, the number of leaders for strong structural controllability are plotted in Figure~\ref{fig:con_comp}. It can be seen that clique chains, which are maximally robust among all graphs with given $N$ and $D$ require many more leaders as compared to the maximally controllable graphs.

\subsection{Robustness Comparison}
\label{sec:robust_sim}
\begin{figure}[h]
\centering
\begin{subfigure}{0.25\textwidth}
\centering
\includegraphics[scale=0.21]{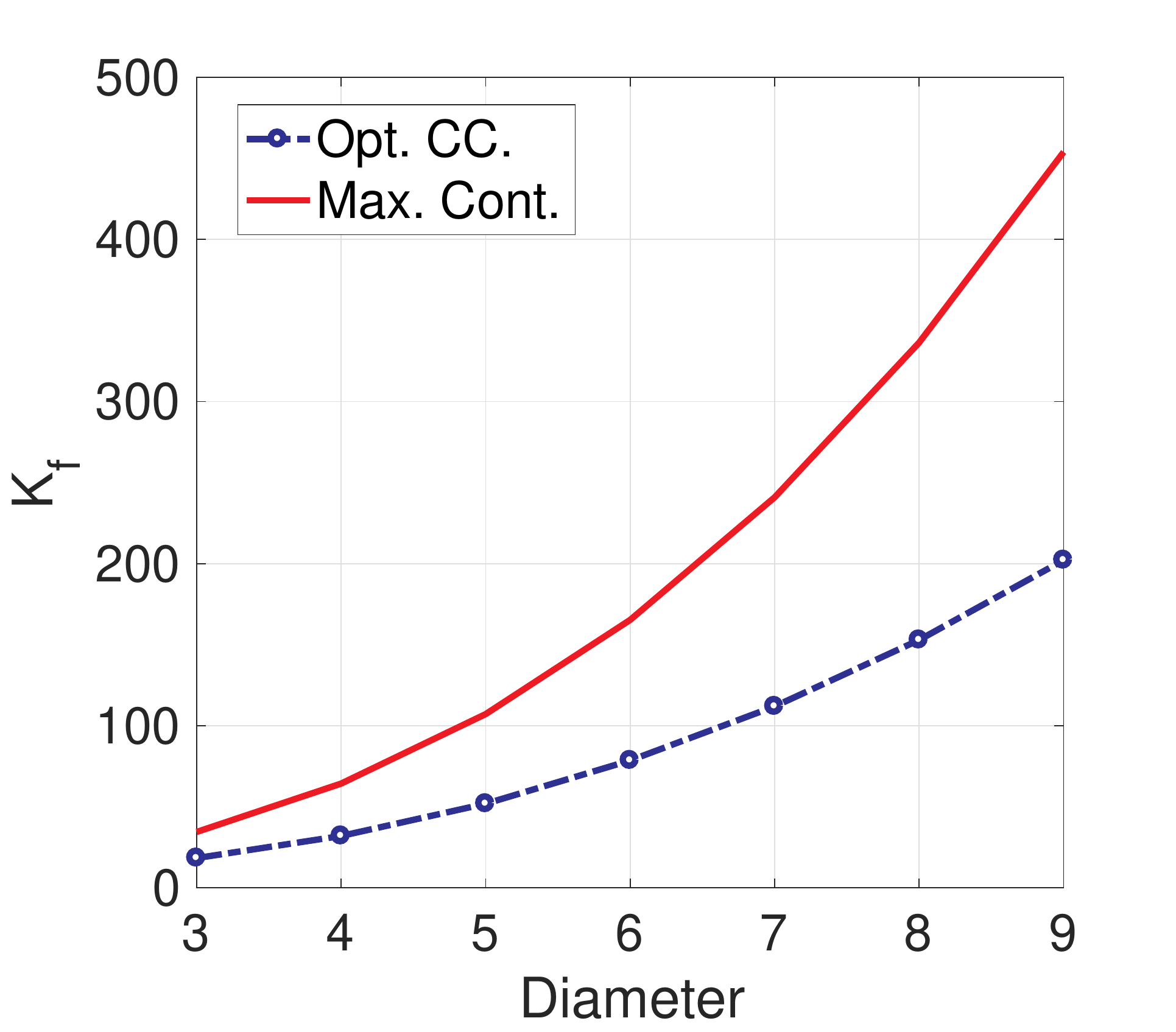}
\caption{$k = 5$}
\end{subfigure}
\begin{subfigure}{0.22\textwidth}
\centering
\includegraphics[scale=0.21]{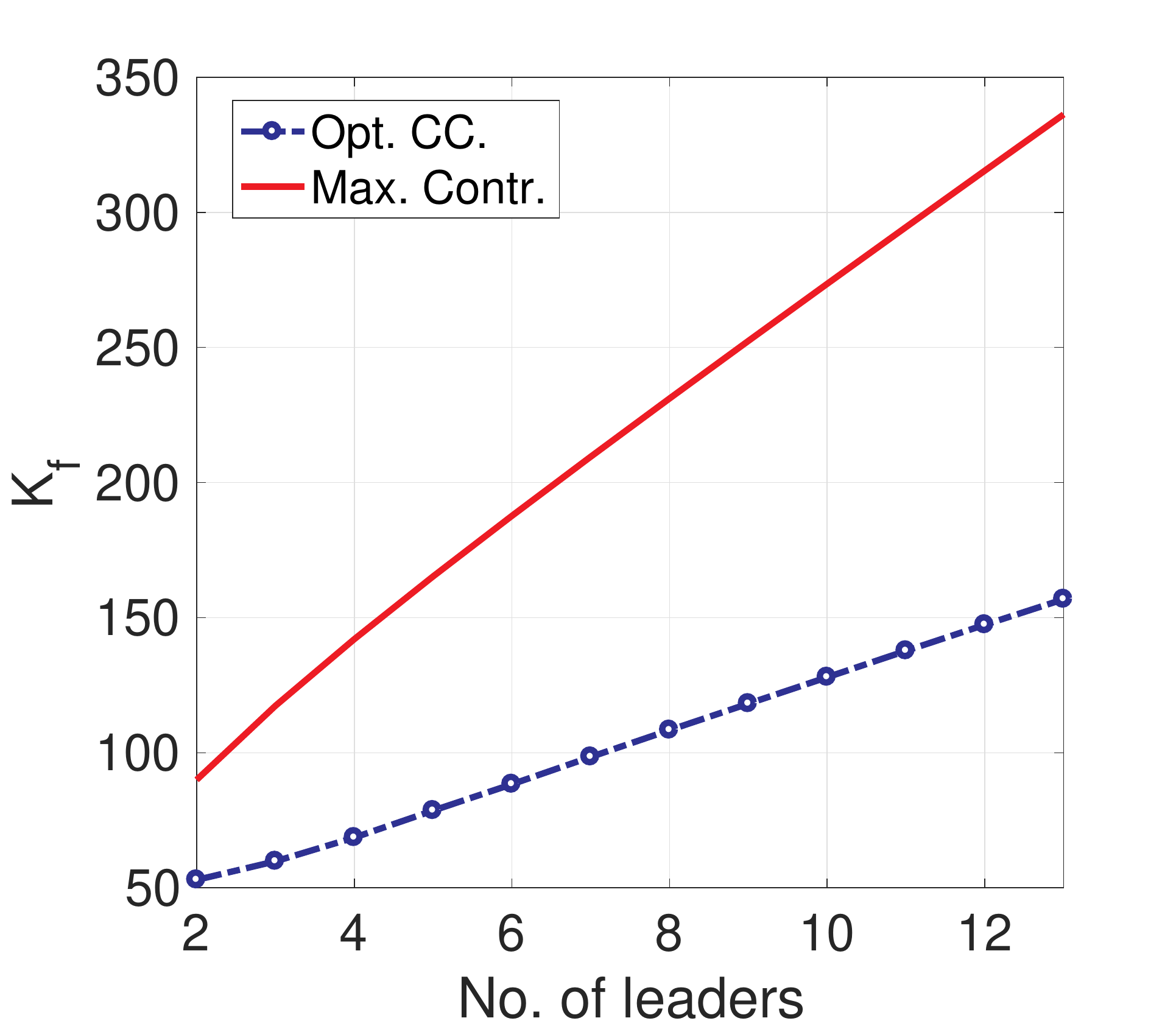}
\caption{$D = 6$}
\end{subfigure}
\caption{Comparison of $K_f$ of the maximally controllable graphs $\bar{\mathcal{M}}$ and the corresponding optimal clique chains with the same $D$ and $N = kD+1$.}
\label{fig:compmyCC}
\end{figure}

In this subsection, we compare Kirchhoff indices of maximally controllable graphs ($\bar{\calM}$) and the corresponding maximally robust graphs for given $N$ and $D$. Although we know that for given $N$ and $D$, maximally robust graphs are clique chains of the form $\mathcal{G}_D(1,n_2,\cdots,n_{D},1)$ where $N=2 + \sum_{i=2}^D n_i$; we do not know the exact values of $n_i$'s in general and compute them numerically. We plot $K_f$ of $\bar{\mathcal{M}}$ and optimal clique chains with the same $D$ and $N$ as a function of $D$ (while fixing $k$) in Figure \ref{fig:compmyCC}(a), and as a function of $k$ (while fixing $D$) in Figure \ref{fig:compmyCC}(b). We observe that $K_f$ of maximally controllable graph is roughly the double of the $K_f$ of the corresponding clique chain, especially for the larger $D$ values. 

In Table \ref{tab:new}, we select $D$ and the number of leaders $k$ and then generate optimal clique chains (through exhaustive search) with $N=kD+1$, and also maximally controllable graphs $\bar{\mathcal{M}}$ (as in Section \ref{sec:MC}) with the same $D$, $k$, and $N$. We again observe that for the same network parameters $D$ and $N$, optimal clique chains are significantly more robust than the corresponding maximally controllable networks.

\begin{table}[]
\centering
\caption{Optimal clique chains and their $K_f$ for a given $D$ and $N$, where $N=kD+1$, along with the $K_f$ of corresponding maximally controllable graphs $\bar{\mathcal{M}}$.}
\begin{tabular}{|l|l|l|l|l|l}
\cline{1-5}
$D$ & $k$ &  $\mathcal{G}^\ast_D(n_1,\cdots,n_{D+1})$ & $K_f(\mathcal{G}^\ast_D)$ & $K_f(\bar{\mathcal{M}})$ &  \\ \cline{1-5}
 3 &  2 & $\left(1\;\;\; 2\;\;\; 3\;\;\; 1\right)$ & 10.5 & 16.64 \\
   &  3 & $\left(1\;\;\; 4\;\;\; 4\;\;\; 1\right)$  & 12.73 & 22.75 \\
 \cline{1-5}
   &  2 & $\left(1\;\;\; 2\;\;\; 3\;\;\;2\;\;\; 1\right)$ & 19.57 & 32.64 \\
4  &  3 & $\left(1\;\;\; 3\;\;\; 5\;\;\;3\;\;\; 1\right)$ & 23.30 & 43.72 \\
   &  4 & $\left(1\;\;\; 4\;\;\; 7\;\;\;4\;\;\; 1\right)$ & 27.59 & 54.19 \\ \cline{1-5}
   &  2 & $\left(1\;\;\; 2\;\;\;2\;\;\; 3\;\;\;2\;\;\; 1\right)$ & 33.75 & 56.56 \\
   &  3 & $\left(1\;\;\; 3\;\;\;4\;\;\; 4\;\;\;3\;\;\; 1\right)$ & 38.73 & 74.63 \\ 
5   &  4 & $\left(1\;\;\; 3\;\;\;6\;\;\; 6\;\;\;4\;\;\; 1\right)$ & 45.32 & 91.27 \\ 
   &  5 & $\left(1\;\;\; 4\;\;\;8\;\;\; 8\;\;\;4\;\;\; 1\right)$ & 51.90 & 107.18 \\   \cline{1-5} 
   &  2 & $\left(1\;\;\; 2\;\;\;2\;\;\; 3\;\;\;2\;\;\; 2\;\;\; 1\right)$ & 52.96 & 89.99 \\
   &  3 & $\left(1\;\;\; 2\;\;\;4\;\;\; 4\;\;\;4\;\;\; 3\;\;\; 1\right)$ & 59.85 & 117.40 \\
6  &  4 & $\left(1\;\;\; 3\;\;\;5\;\;\; 6\;\;\;6\;\;\; 3\;\;\; 1\right)$ & 68.62 & 142.04 \\
   &  5 & $\left(1\;\;\; 4\;\;\;7\;\;\; 7\;\;\;7\;\;\; 4\;\;\; 1\right)$ & 78.62 & 165.27 \\
   &  6 & $\left(1\;\;\; 4\;\;\;8\;\;\; 10\;\;\;9\;\;\; 4\;\;\; 1\right)$ & 88.36 & 187.67 \\  \cline{1-5}
   &  2 & $\left(1\;\;\; 2\;\;\;2\;\;\; 2\;\;\;3\;\;\; 2\;\;\; 2\;\;\;1\right)$ & 79.24 & 134.54 \\
   &  3 & $\left(1\;\;\; 2\;\;\;4\;\;\; 4\;\;\;4\;\;\; 4\;\;\;2\;\;\; 1\right)$ & 86.42 & 173.96 \\
 &  4 & $\left(1\;\;\; 3\;\;\;5\;\;\; 5\;\;\;6\;\;\; 5\;\;\;3\;\;\; 1\right)$ & 98.61 & 208.65 \\
7 &  5 & $\left(1\;\;\; 3\;\;\;6\;\;\; 8\;\;\;8\;\;\; 6\;\;\;3\;\;\; 1\right)$ & 111.94 & 240.89 \\
  &  6 & $\left(1\;\;\; 4\;\;\;7\;\;\; 9\;\;\;9\;\;\; 8\;\;\;4\;\;\; 1\right)$ & 125.64 & 271.72 \\
  &  7 & $\left(1\;\;\; 4\;\;\;9\;\;\; 11\;\;\;11\;\;\; 9\;\;\;4\;\;\; 1\right)$ & 139.37 & 301.66 \\ \cline{1-5}
\end{tabular}
\label{tab:new}
\end{table}

\section{Conclusions}
\label{sec:con}
Networks that exhibit higher robustness to noise and structural changes typically require many leader nodes (inputs) to become controllable. For a fixed number of nodes $N$, complete graphs are maximally robust but require $(N-1)$ leaders for complete controllability. At the same time, path graphs require only one leader for controllability; however, such graphs are minimally robust. We observed a similar relationship between controllability and robustness if we also fix the diameter $D$ of a graph along with the number of nodes $N$. Clique chains are optimal from the robustness perspective for given $N$ and $D$. However, they require a large number of leaders, either $N-(D+1)$ or $N-D$, for strong structural controllability. On the other hand, for arbitrary $N$ and $D$, we can construct graphs that are strong structurally controllable with at most $\lceil\frac{N-1}{D}\rceil$ leaders, which is a sharp upper bound. However, such graphs are much less robust than optimal clique chains with the same $ N $ and $ D $. Graph-theoretic tools for network controllability, including equitable partitions and distances of nodes to leaders, are particularly useful to exploit the controllability and robustness trade-off. In the future, we aim to explore graph operations that maximally improve one of the two properties while minimally deteriorating the other. 

\section*{Acknowledgment}
We thank anonymous reviewers for their comments that improved the quality of the paper.

\bibliographystyle{IEEEtran}
\bibliography{references} 

\end{document}